\documentclass[preprintnumbers,aps,twocolumn,floatfix,superscriptaddress]{revtex4-1}
\usepackage{eso-pic,calc}
\usepackage{graphicx}
\usepackage{amsmath, amsthm, amssymb}
\usepackage{epsfig}
\usepackage{bm}
\usepackage{color}
\usepackage{bbding}
\usepackage{wasysym}
\usepackage{enumitem}
\usepackage{booktabs}

 \usepackage[colorlinks]{hyperref}
\hypersetup{citecolor=blue}

\def\ie{i.e.~}
\def\etl{$et~al.$~}

\def\beqr{\begin{eqnarray}}
	\def\eqnr{\end{eqnarray}}
\def\beq{\begin{equation}}
	\def\bc{\begin{center}}
		\def\ec{\end{center}}
	\def\eqn{\end{equation}}

\begin{document}

\title{Diversity mitigates polarization and consensus in opinion dynamics}
	
\author{Sidharth Pradhan}
\affiliation{Department of Physics, Institute of Science, Banaras Hindu University, Varanasi, Uttar Pradesh 221 005, India}
	
\author{Sangeeta Rani Ujjwal}
\affiliation{Department of Physics, Institute of Science, Banaras Hindu University, Varanasi, Uttar Pradesh 221 005, India}

\begin{abstract}
{We study the opinion dynamics in a population by considering a variant of the Kuramoto model where the phase of an oscillator represents the opinion of an individual on a single topic. Two extreme phases separated by $\pi$ represent opposing views. Any other phase is considered as an intermediate opinion between the two extremes. The interaction (or attitude) between two individuals depends on the difference between their opinions and can be positive (attractive) or negative (repulsive) based on the defined thresholds. We investigate the opinion dynamics when these thresholds are varied. We observe an explosive transition from a bipolarised state to a consensus state with the existence of scattered and tri-polarised states at low values of threshold parameter. The system exhibits multistability between various states in a sizable parameter region. These transitions and multistability are studied in populations with different degrees of diversity represented by the width of conviction distribution. We found that a more homogeneous population has greater tendency to exhibit bipolarised, tri-polarised and clustered states while a diverse population helps mitigate polarisation among individuals by reaching a consensus sooner. Ott-Antonsen analysis is used to analyse the system's long term macroscopic behaviour and verify the numerical results. We also study the effect of neutral individuals that do not interact with others or do not change their attitude on opinion formation, nature of transitions and multistability. Furthermore, we apply our model to language data to study the assimilation of diverse languages in India and compare the results with those obtained from model equations.   }
\end{abstract}

\maketitle

\section{Introduction}

Polarisation of perspectives in a crowd can either be adverse~\cite{axelrod2021, wu2023} or beneficial~\cite{tran2010} depending upon the matter under consideration. It may cause a conflicted group to lose collective advantage, such as failing to unite even for a common cause~\cite{macy2021} or it can help preserve cultures within cohesive, consensus-seeking groups. Transitions between collective states of a group of individuals have been studied and tipping points in the opinion dynamics have been identified in some past studies~\cite{galam2020,macy2021}. Such transitions can be reversible or irreversible~\cite{macy2021,nettasinghe2025}, which open avenues for strategic manipulation of states, for instance averting polarisation among individuals~\cite{flache2017}. The collective behavior of a crowd is often determined by interactions based on the opinions of the individuals~\cite{warren2024}. These opinion oriented interactions between pairs of agents can either induce an assimilation effect leading to a consensus or a contrast effect~\cite{sherif1958} resulting in scattered opinions. Such interactions can also lead to formation of opinion clusters~\cite{warren2024}. One theory that deals with interactions based on opinions is the social judgement theory (SJT)~\cite{hunter1984}. SJT is a persuasion theory~\cite{okeefe2006} which places opinions on an attitude scale~\cite{sherif1965} that leads to attitude formation and opinion change of an individual based on the opinion of the other individual it is interacting with. According to this theory, an individual uses his present opinion to categorize other opinions into three zones on the attitude scale~\cite{jager2005}: the latitude of acceptance, the latitude of non-commitment, and the latitude of rejection, in increasing order of difference of one's opinion with the opinion of other individual (s)he is interacting with. These categories shape an individual's attitude towards different opinions and may result in a shift in one's own stance.

Like many real life systems, opinion dynamics in a social setup can be studied using a network science framework where the system can be considered as a network consisting of nodes and edges wherein the nodes represent individuals in the system and the edges denote interactions between them. Interactions between these nodes can lead to rich and sometimes astonishing dynamics in the system. Real life systems consisting of discrete, interacting subsystems can be effectively treated as a system of coupled oscillators~\cite{iatsenko2013}. In this context, the Kuramoto model (KM)~\cite{acebron2005, coutinho2013} has proved to be a simple yet effective model to study the collective behaviours of such systems. The KM and its variants have been used to understand the dynamics of various natural and artificial systems including brain~\cite{maistrenko2007, sheeba2008}, Josephson junction array~\cite{daniels2003, trees2005}, neutrino flavour oscillations~\cite{pantaleone1998} and opinion depolarisation~\cite{ojer2023}. The classical KM incorporates the pairwise symmetric attractive coupling between the oscillators and captures the spontaneous transition from desynchronized state to synchronized dynamics~\cite{acebron2005} as the coupling strength increases. In a network system when the coupling between the interacting oscillators is both attractive and repulsive, $\pi$ states~\cite{hong2011} and traveling wave states~\cite{iatsenko2013} have been observed. On the other hand when a particular oscillator is assigned heterogenous couplings with asymmetric pairwise interaction~\cite{hong2012}, $\pi$ and travelling wave states do not appear. Symmetric and asymmetric interactions show evidence of multiple clustered states as reported in a study with m\"obius strip~\cite{ren2013}. Further, models of opinion dynamics incorporating homophily~\cite{bessi2016} and echo chambers~\cite{baumann2020} also show results aligning with polarised and consensus states. Taking into account the importance of relating and integrating different models~\cite{flache2017, li2023}, we aim to study dynamics of opinions on a single topic within a group using a variant of KM in conjunction with the SJT. Attractive and repulsive couplings in the model represent social attitudes, which help in identifying tipping points that can potentially be used to manage or mitigate polarisation and other undesirable states.

In this work, we consider a model inspired by the Kuramoto model to study opinion dynamics when the interactions between different individuals depend upon their opinions. In this model the opinion of an individual on a particular topic is represented by a phase variable. The two extreme opinions on a topic have a phase difference of $\pi$ on a cyclic phase scale. Any intermediate opinion between them is characterised by phase lying between these extremes. We define two thresholds $A$ and $B$ as limiters within range $[0,\pi]$ to assign the nature of interactions (couplings) between the two individuals on the basis of difference between their opinions. If the difference is less than $A$ the interaction is attractive whereas when the difference is more than $B$, the interaction between the individuals is repulsive. The differences lying between $A$ and $B$ falls in the neutral region where the individuals either do not interact or do not change their previously attained attitudes. The pairwise interactions are considered to be symmetric and the natural frequency drawn from a distribution represents the conviction or strength of stance of an individual towards a specific topic. We study the collective dynamics of the opinions of interacting individuals by changing the parameter $A$ (or $B$). Increasing $A$ effectively enlarges the latitude of acceptance and reduces the latitude of rejection, thereby increasing the fraction of attractively coupled oscillator pairs. We observe different dynamical states such as scattered, tri-polarised,  bipolarised ($\pi$-state) and consensus on increasing the range of attractive coupling, $A$. Increasing the parameter $A$ and on relaying the phases of oscillators from the previous run to the next value of parameter, latency in increase in attractive attitudes with respect to limiters ($A$ and $B$) is observed along with explosive transition~\cite{boccaletti2016} from $\pi$ state to consensus giving rise to tipping points and hysteresis loop in the system. Also two step transition with the occurrence of tri-polarised states is observed. The numerically obtained results are validated using the Ott-Antonsen analysis~\cite{ott2008, ott2009}. The relevance of the findings of this model is checked by applying the model on the empirical data of language evolution and assimilation. Our study shows how different attitudes and their ranges influence the assimilation processes of languages and its reversibility.

The paper is arranged as follows. Sec. II describes the model used to study opinion dynamics. The emerging dynamical states and transitions are discussed in Sec. III. Dimensional reduction using Ott-Antonsen formalism and comparison with the numerical results is presented in Sec. IV. In Sec. V the effect of neutral region on the collective opinions of the population is studied. Sec. VI discusses the implications of our model in the real data of language assimilation.  In Sec. VII the population states and tendencies are shown in the coupling parameter space. The results are summarised in Sec. VIII.

  \section{Model}
  
\begin{figure}[h]

   % \begin{minipage}[t]{0.18\textwidth} % Adjusted width
       \hspace*{-1.0cm}
        \includegraphics[scale=0.32]{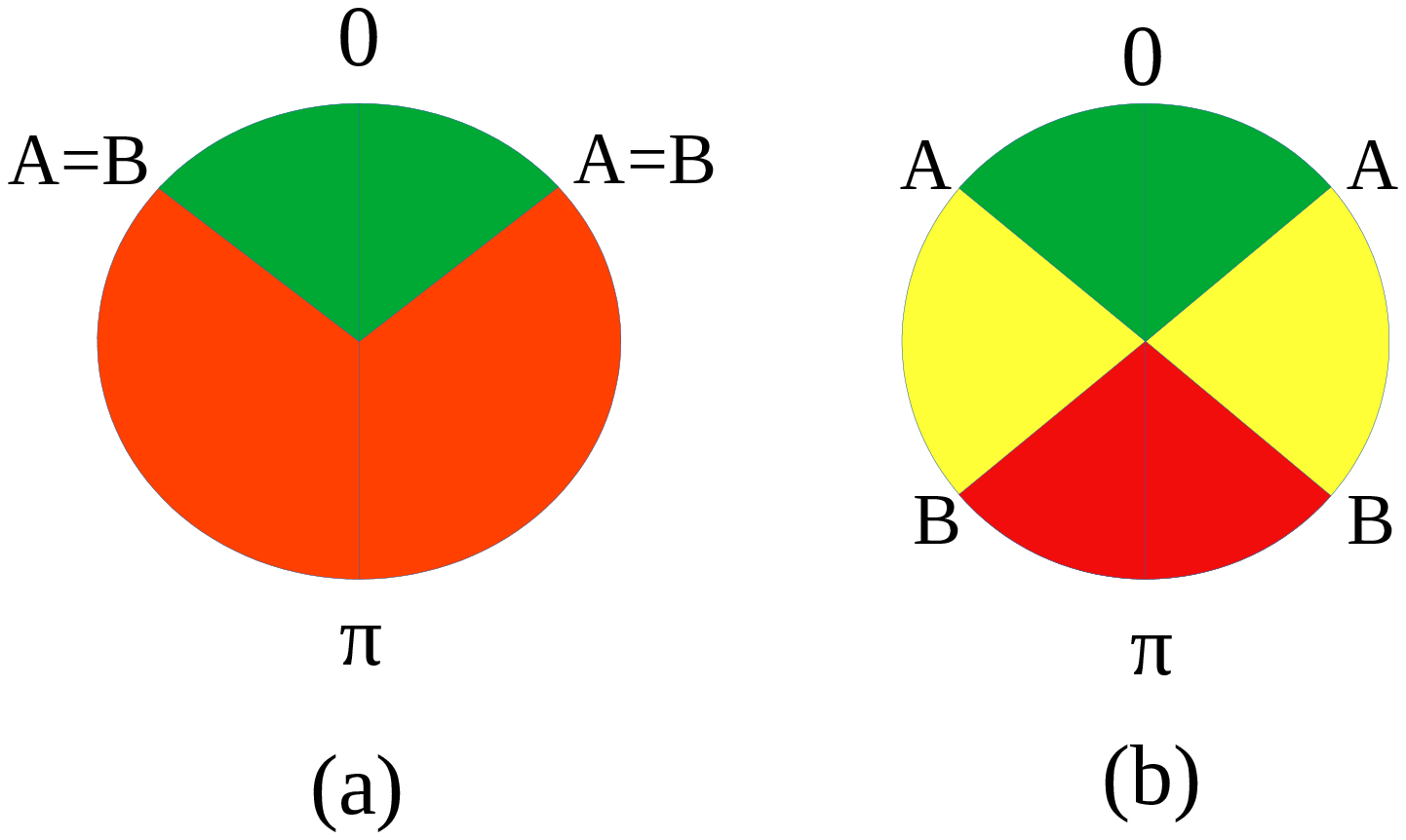}
       \caption{Pictorial representation of limiters $A$ and $B$ for assigning attractive and repulsive couplings (a) without and (b) with the neutral region on the phase cycle.}
       \label{fig:fig1}
   % \end{minipage}
\end{figure}

We model the opinion of an individual on a particular topic as a phase of an oscillator. The opinions can have values in the interval $[0,2\pi]$ and these opinions change with time according to the conviction of an individual represented by the natural frequency of that oscillator, and the nature of interaction between the individuals. The model equation reads as: 
  \begin{equation}
     \frac{d\theta_i}{dt}=\omega_i + \frac{1}{N} \sum_{j=1}^N K_{ij}  \sin(\theta_j - \theta_i),
     \label{eq:eq1}
  \end{equation}
  
where $\theta_i$ and $\omega_i$ denote the opinion and the conviction of the $i^{th}$ individual respectively. $K_{ij}$ is the element of the coupling matrix that is related to the attitude of $i^{th}$ individual towards the $j^{th}$ individual and can take values that can be positive, negative or zero. In the model the coupling matrix $K$ is symmetric i.e $K_{ij}$=$K_{ji}$. The values to $K_{ij}$ are assigned on the basis of the difference in opinions, $|\theta_j - \theta_i|$. Incorporating ideas from SJT, we define two thresholds: $A$ and $B$ on a scale from 0 to $\pi$ (maximum possible phase difference) (Fig. ~\ref{fig:fig1}) and the coupling strength (attitude) for a pairwise interaction is provided according to the following rule:

\begin{equation}
	K_{ij} =
	\begin{cases}
		K_1 & \text{if, } 0 < |\theta_j -\theta_i| \leq A \\
		0 & \text{if, } A < |\theta_j -\theta_i| \leq B \\
		K_2  & \text{if, } B < |\theta_j -\theta_i| \leq \pi .
	\end{cases}
	\label{eq:eq2}
  \end{equation} 
 
Here $K_1$ and $K_2$ are the strengths of attractive and repulsive interactions respectively. $K_{ij}$'s are time-independent and are updated when there is change in the threshold parameters $A$ or $B$ according to the rule given in Eq.~\ref{eq:eq2}. The ratio of negative to positive coupling is defined as $Q=-K_2/K_1$.  The conviction of individuals, $\omega$'s are generated from the unimodal Lorentzian probability distribution \begin{equation}
    g(\omega)=\gamma/[\pi(\omega^2 + \gamma^2)]
    \label{eq:eq3}
\end{equation} with width $\gamma$. 

To characterize the degree of coherence in the collective dynamics of opinions in the population, we define the complex order parameter
  \begin{equation}
      Z=Re^{i\phi}=\frac{1}{N} \sum_{j=1}^N e^{i\theta_j}
      \label{eq:eq4}
  \end{equation}
  and the weighted order parameter of the $i^{th}$ oscillator as:
\begin{equation}
      W_i=S_ie^{i\Phi_i}=\frac{1}{N} \sum_{j=1}^N K_{ij} e^{i\theta_j}.
      \label{eq:eq5}
  \end{equation}  
  
Here, $R$ and $\phi$ are the amplitude of the order parameter and the average phase respectively. $S_i$ is the amplitude and $\Phi_i$ is the average phase of the weighted order parameter $W_i$ of the $i^{th}$ individual. $R$ and $S_i$ range from 0 to 1. The quantity $R$ can be understood as a macroscopic mean field created by the oscillators and measures the collective coherence or ``agreement" among individuals in a population.

We explore different dynamical states arising in the model as $A$ varies. For a fixed value of $A$ and $B$, we start with random initial $\theta \in [0,2\pi]$ and assign $K_{ij}$ according to the rule given in Eq. ~\eqref{eq:eq2} by computing the pairwise differences, $(\theta_j -\theta_i)$ for all pairs. The system is evolved to reach the asymptotic stable state and the order parameter, $R$ is computed. Note that on the cyclic phase scale the maximum distance between two phases is $\pi$. The phase difference between an interacting pair determines their attitude based on three defined regions as depicted in Fig.~\ref{fig:fig1}(b):
\begin{itemize}

 \item   If it lies in the green region (less than $A$), the interaction falls within the latitude of acceptance (attractive coupling)

 \item   If it lies in the yellow region (between $A$ and $B$), it is classified as latitude of non-commitment (neutral coupling)

 \item   If it lies in the red region (greater than $B$), it belongs to the latitude of rejection (repulsive coupling)
\end{itemize}   

This framework allows us to investigate how changes in the range of accepted or rejected differences influence the opinion dynamics and consensus formation which will be discussed in the following section.

\section{States and transitions}
We begin by considering the case where the latitude of non-commitment is absent meaning $A$ and $B$ coincide ($B-A=0$) as shown in Fig.~\ref{fig:fig1}(a). In this configuration, the possible attitudes or couplings between the individuals are either attractive or repulsive in nature. When $A=0$, all differences $|\theta_j - \theta_i|$ fall into the latitude of rejection, leading to purely repulsive interactions between the individuals. On the other hand, when $A=\pi$, which is the largest possible distance, all opinion differences will lie in the latitude of acceptance. The former case i.e $A=0$ results in a scattered state where the opinions of all the individuals are different, spreading across the entire range between 0 to $2\pi$ while the later case i.e $A=\pi$ leads to a consensus where the opinions of all individuals are clustered around the same value.

\begin{table}[htbp]
\centering
\caption{System parameters used for simulations.}
\begin{tabular}{lll}
\toprule
\textbf{Parameter} & \textbf{Description} & \textbf{Numerical Value} \\
\midrule
$N$ & Number of agents & $500$ \\
$K_1$ & Attractive coupling strength & $1.0$ \\
$K_2$ & Repulsive coupling strength & $-0.5$ \\
$Q$ & $Q=-K_2/K_1$ & $0.5$ \\
$\delta t$ & Time step & $0.01$ \\
$n_{steps}$ & Transient time steps & $2\times10^5$ \\
$\theta_i$ & Initial opinions & $\theta_i \in [0, 2\pi]$ \\
$g(\omega)$ & Conviction distribution & $ \frac{\gamma}{\pi(\omega^2 + \gamma^2)}$ \\
$\delta A, \delta B$ & Small change in $A$, $B$ & 0.01$\pi$\\

\bottomrule
\end{tabular}
\end{table} 

\begin{figure}[h]
       
        \centering
        \hspace{-0.2cm}
        \includegraphics[width=0.5\textwidth]{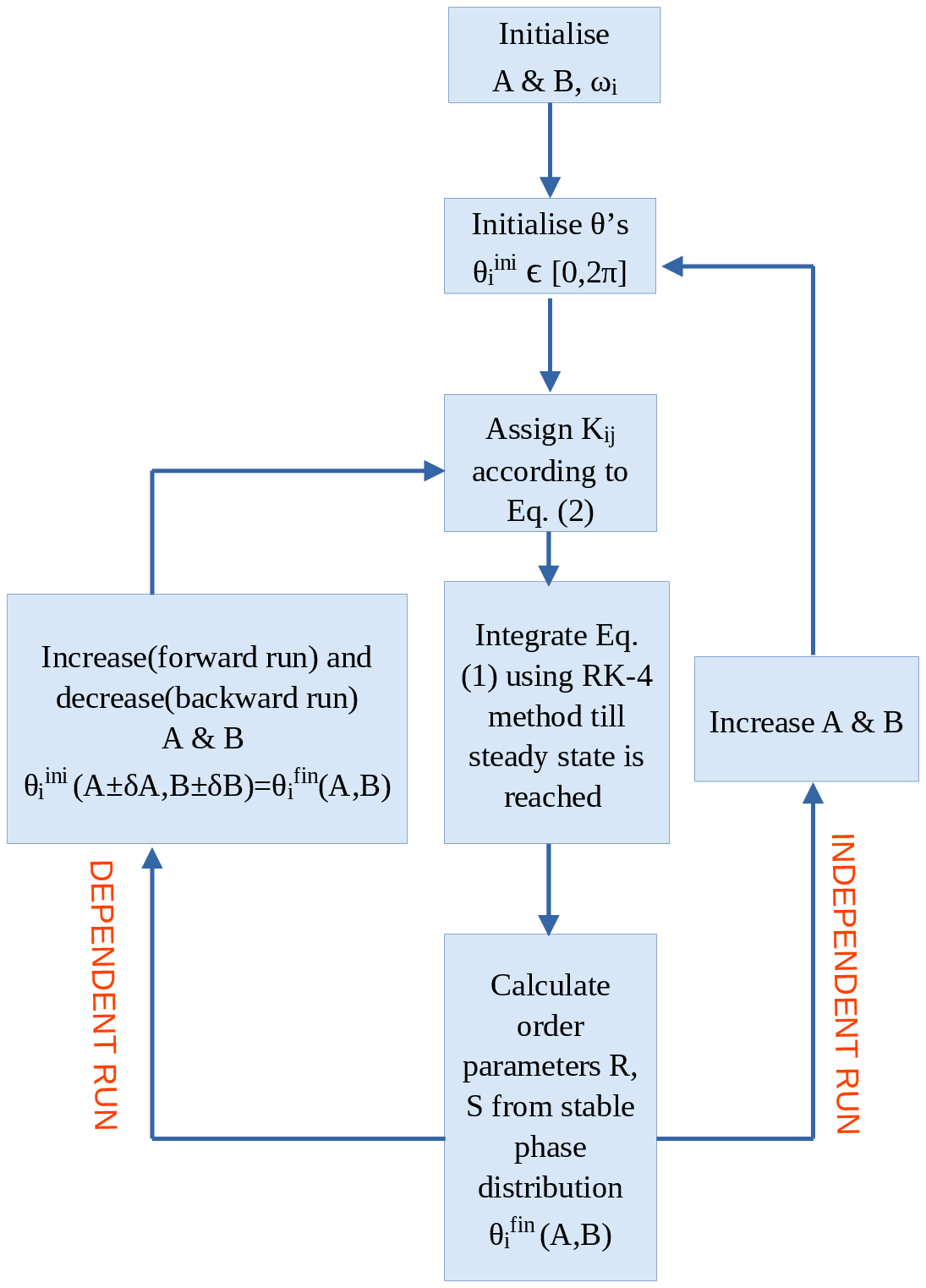} % First image
    
    \caption{Block diagram outlining the steps followed to calculate order parameters numerically from Eqs.~\eqref{eq:eq1}-\eqref{eq:eq4} for dependent and independent runs. }
    \label{fig:fig2}
\end{figure}

\begin{figure}[h]
       
        \centering
        \hspace{-0.2cm}
        \includegraphics[width=0.48\textwidth]{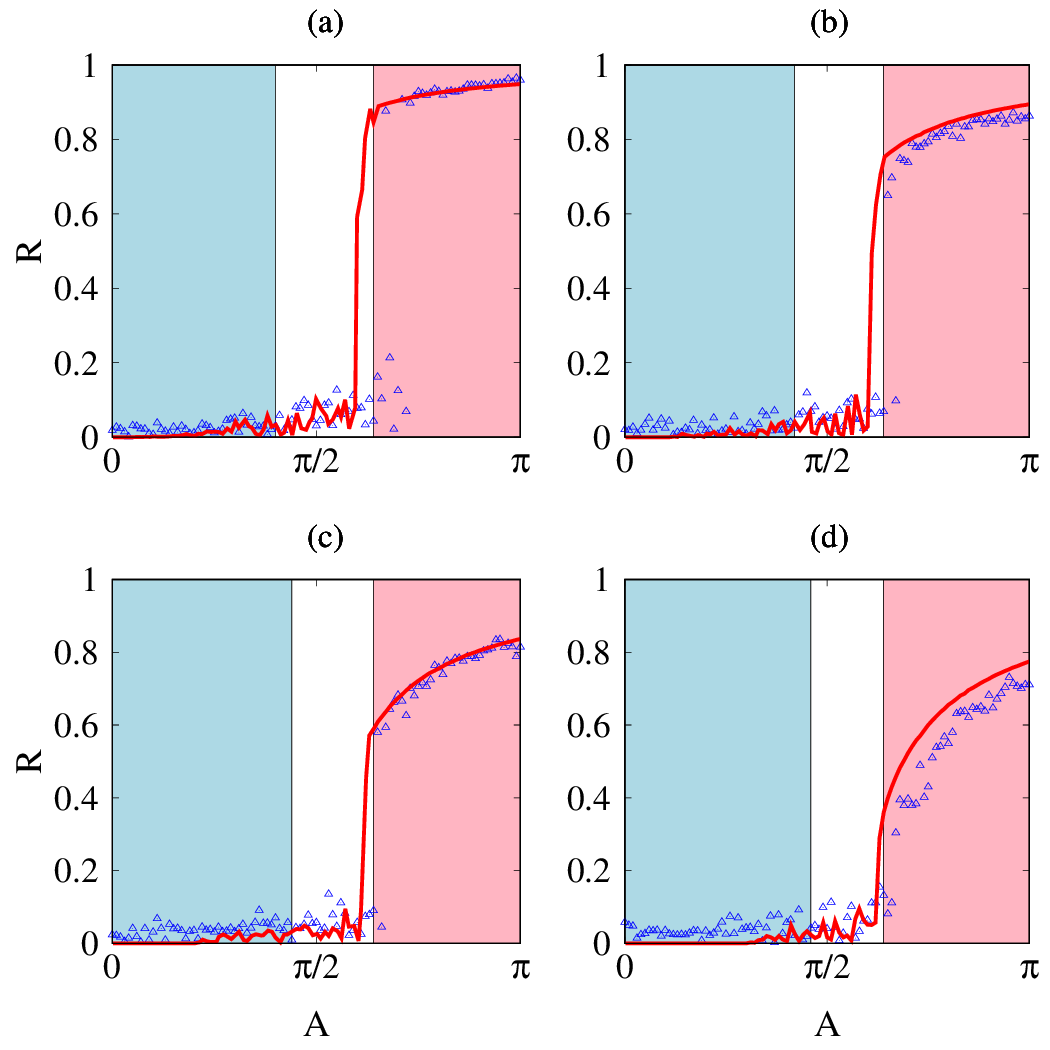} % First image
    
    \caption{ Variation of order parameter $R$ with increase in $A$ without neutral region ($A=B$) shown for different conviction width, (a) $\gamma$=0.05, (b) $\gamma$=0.10, (c) $\gamma$=0.15, and (d) $\gamma$=0.20. The numerical results obtained by integrating Eq.~\eqref{eq:eq1} are plotted with blue triangles while the analytical predictions from Ott-Antonsen analysis (from Eqs.~\eqref{eq:eq19}-~\eqref{eq:eq24} ) are plotted with red solid lines.}
    \label{fig:fig3}
\end{figure}

\begin{figure}[h]

        \centering
        \hspace{-0.2cm}
        \includegraphics[width=0.48\textwidth]{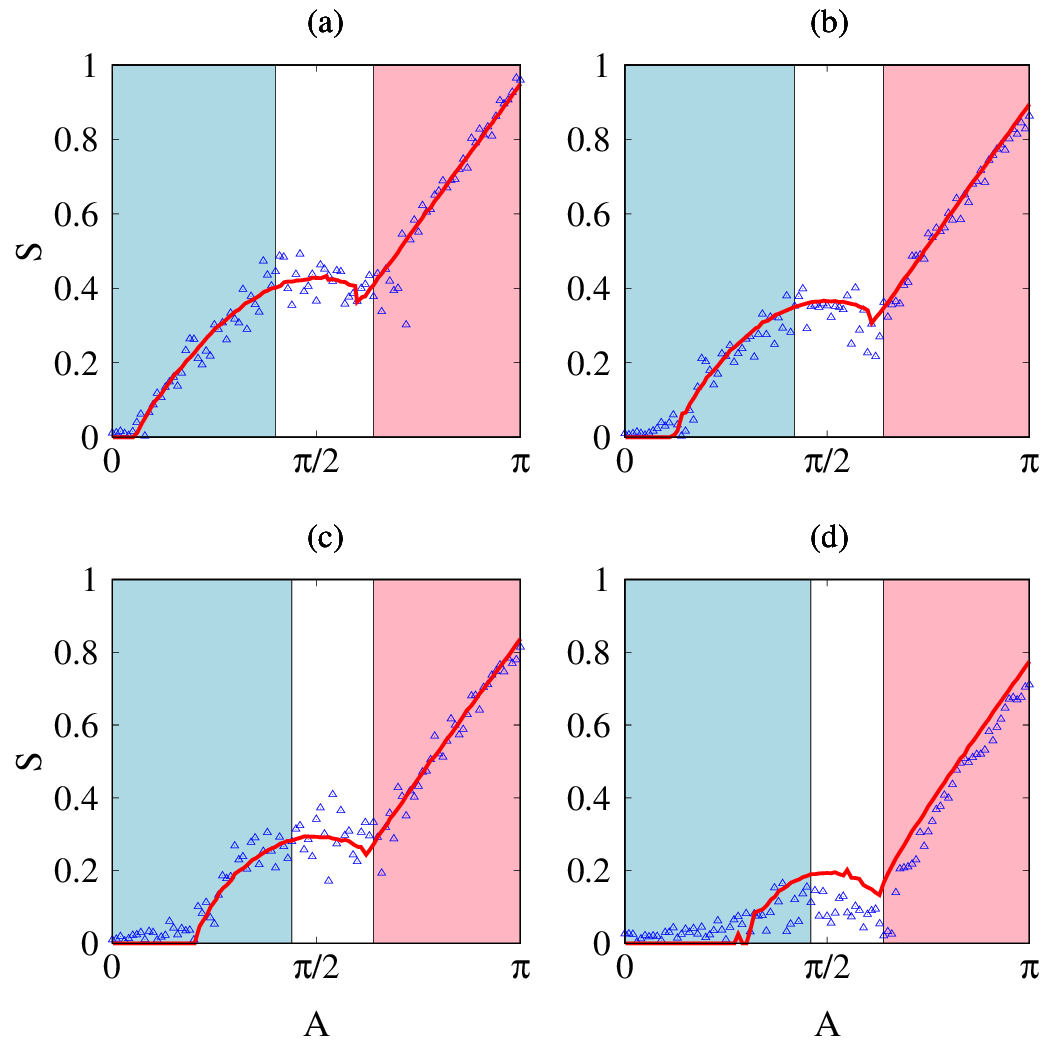} % First image
    
    \caption{ Average weighted order parameter, $S$ (Eq.~\eqref{eq:eq25}) is plotted as a function of $A$ for the case when there is no neutral region ($A=B$) for different values of conviction spread: (a) $\gamma$=0.05, (b) $\gamma$=0.1, (c) $\gamma$=0.15, and (d) $\gamma$=0.2. The results obtained by numerically integrating Eq.~\eqref{eq:eq1} are shown with blue triangles and the theoretical predictions are shown with red solid lines (from Eqs.~\eqref{eq:eq19}-\eqref{eq:eq25}).}
    \label{fig:fig4}
\end{figure}

We integrate Eq.(~\ref{eq:eq1}) using the fourth order Runge-Kutta method at a fixed value of $A$. The parameters used for simulations are provided in Table 1. These parameters remain the same for all simulation results presented in the paper unless stated otherwise. The order parameter $R$ is computed for each $A$ after the system reaches steady state. To get $R$ vs $A$ plot, the threshold $A$ is increased from 0 to $\pi$ in steps of $0.01\pi$. The steps involved in the simulation for independent and dependent runs are outlined in the block diagram shown in Fig.~\ref{fig:fig2}. First we present the results for the independent runs where $\theta_i$ are taken uniformly from $ [0, 2\pi]$ at each value of $A$. We observe from Fig.~\ref{fig:fig3} that at low $A$, the system exhibits scattered state. As $A$ increases the population becomes polarised forming two groups with diagonally opposite opinions on phase scale. The population forms a consensus as $A$ approaches the maximum value \ie  $\pi$. The transition from $\pi$ state to the consensus state is abrupt and investigated by examining the distribution of the order parameter at the transition point which displays a clear bimodality and Binder cumulant ~\cite{binder1981, binder1984} which exhibits a pronounced negative dip providing evidences that this transition is similar to a first-order phase transition. In order to investigate the effects of conviction of individuals in the population on the emerging dynamical states and state transitions, the variation of $R$ with $A$ is plotted for a different width $\gamma$ of the distribution $g(\omega)$ (see Fig.~\ref{fig:fig3}). In Fig.~\ref{fig:fig3}, the numerically obtained variation of order parameter, $R$ on varying $A$ is shown with blue triangles. The light blue, white and pink background colors in Fig.~\ref{fig:fig3} denote the regions of scattered, $\pi$ and consensus states respectively. As the width $\gamma$ increases the value of $R$ in the consensus region decreases (Fig.~\ref{fig:fig3}(a)-(d)). This observation suggests that when there is more spread in the conviction of people it is difficult to get a high consensus state (characterised by $R$=1) due to the presence of inflexible individuals in a population as reported earlier ~\cite{galam2007}. This is because the distribution with zero mean and lesser width implies a population with overall less conviction, meaning more susceptible to change in their opinions and hence forming a consensus. Another interesting observation here is that as the conviction width $\gamma$ increases the region of existence of $\pi$ or bipolarised state decreases, indicating that a more diverse population will have less tendency of getting polarised~\cite{han2019, shirzadi2024}. Since the simulation for each value of $A$ is initialized with random phases with each run independent of each other, we refer to these simulations as independent runs.
In this case the percentage of attractively coupled oscillators increases linearly with increase in $A$. The variation of the average weighted order parameter, $S$ with $A$ can be seen in Fig.~\ref{fig:fig4}.  

  \begin{figure}[h]
  
        \centering
        \hspace{-0.2cm}
        \includegraphics[width=0.48\textwidth]{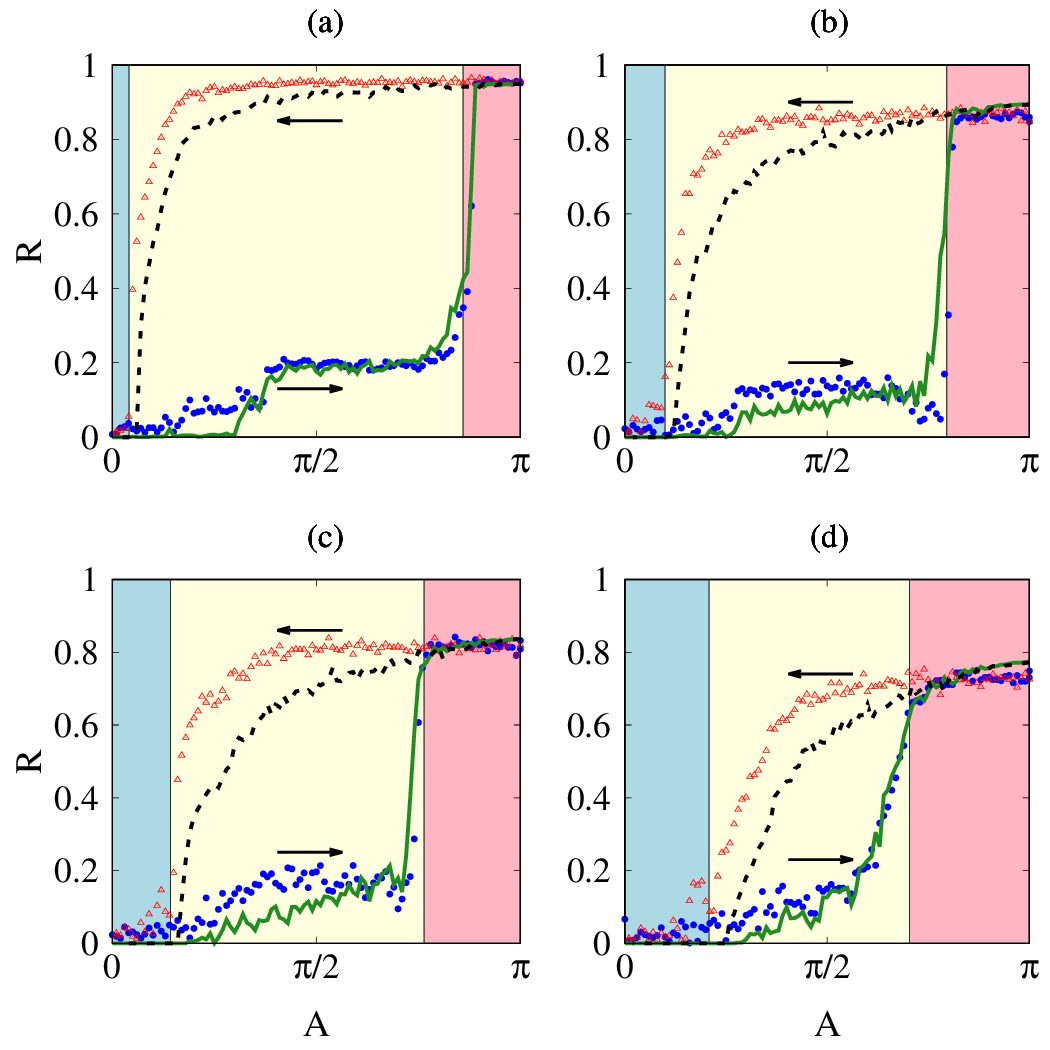} % First image
    
    \caption{Order parameter, $R$ is plotted with varying $A$ ($A=B$) in the forward and backward directions for the dependent runs at different values of distribution width: (a) $\gamma$=0.05, (b) $\gamma$=0.10, (c) $\gamma$=0.15, and (d) $\gamma$=0.20. The numerical results obtained from Eqs.~\eqref{eq:eq1}-~\eqref{eq:eq4} are shown with blue circles (forward run) and red triangles (backward run). The theoretical values (from Eqs.~\eqref{eq:eq19}-~\eqref{eq:eq24} ) are shown with green solid lines (forward run) and black dashed lines (backward run).}
     \label{fig:fig5}
\end{figure}

\begin{figure}[h]
   
        \centering
        \hspace{-0.2cm}
        \includegraphics[width=0.48\textwidth]{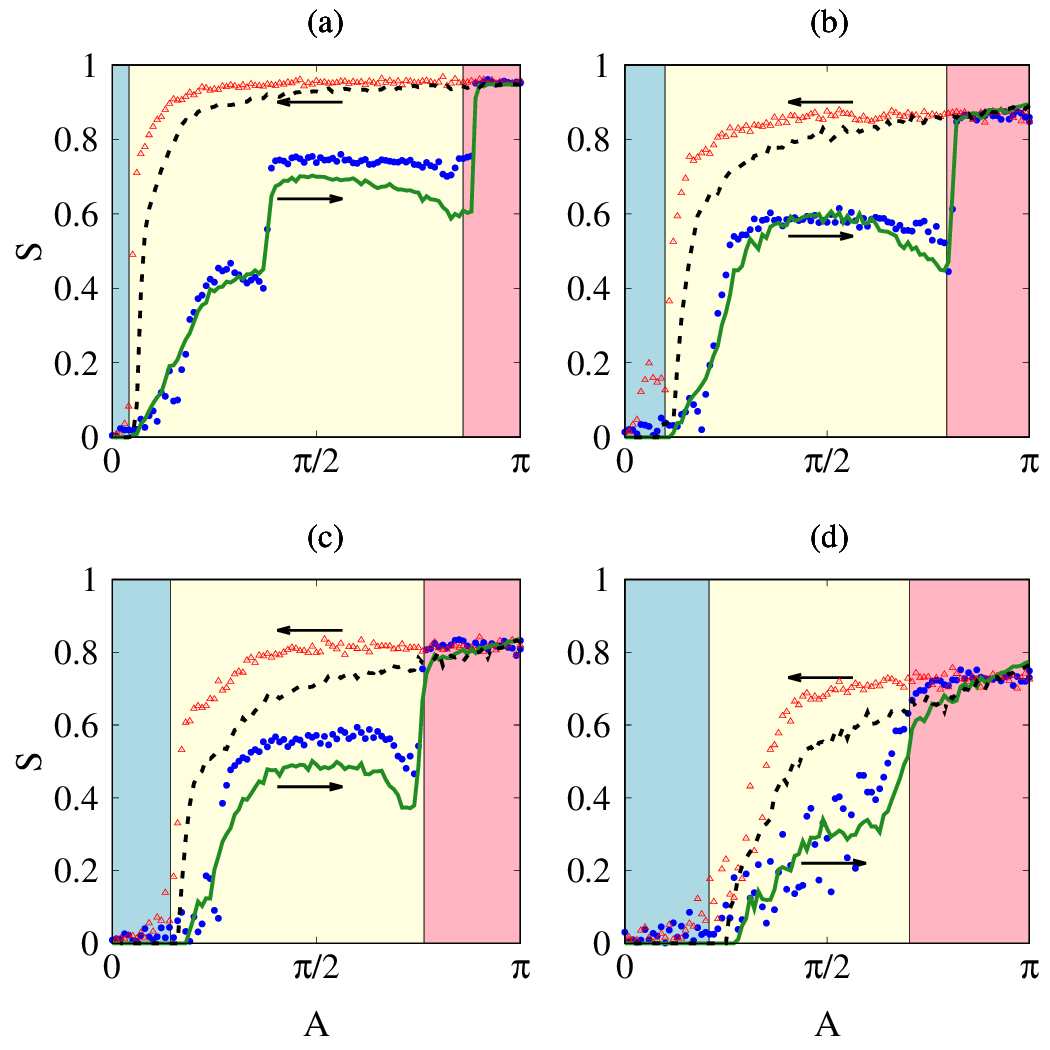} % First image
    
    \caption{Average weighted order parameter, $S$ is plotted with varying $A$ ($A=B$) in the forward and backward directions for the dependent runs at different values of distribution width: (a) $\gamma$=0.05, (b) $\gamma$=0.10, (c) $\gamma$=0.15, and (d) $\gamma$=0.20. The numerical results obtained from Eqs.~\eqref{eq:eq1}-~\eqref{eq:eq5} are shown with blue circles(forward run) and red triangles(backward run). The theoretical values (from Eqs.~\eqref{eq:eq19}-~\eqref{eq:eq25} ) are shown with green solid lines (forward run) and black dashed lines (backward run).}
     \label{fig:fig6}
\end{figure}

\begin{figure}[h!]
        \centering
        \hspace{-0.2cm}
        \includegraphics[width=0.48\textwidth]{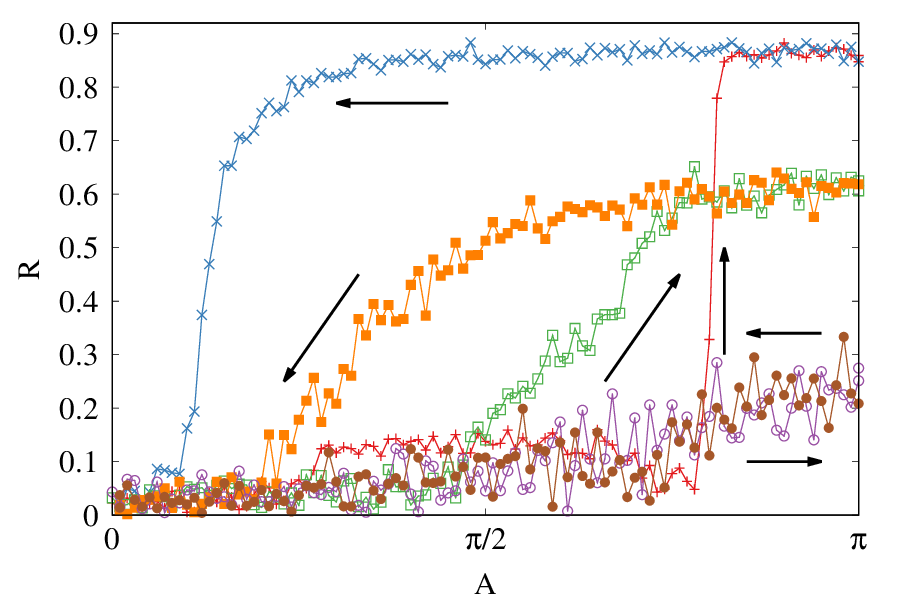} % First image
    
    \caption{Order parameter, $R$ is plotted with varying $A$ ($A=B$) in the forward and backward directions for the dependent runs at different values of distribution width: $\gamma=0.1$ [forward(red + ), backward (blue $\times$)], $\gamma=0.3$ [forward (green open squares), backward (orange filled squares)] and $\gamma=0.5$ [forward (purple open circles), backward(brown filled circles)].}
    \label{fig:fig7}
\end{figure}

One of the central aspects of SJT is changing one's own attitude by placing other's opinions on an attitude scale and comparing them with one's present state of opinion~\cite{hunter1984, sherif1965, jager2005}. In addition, considering the exchange of opinions as a fast process and the change in the range of interaction as a slow process ~\cite{starnini2025}, interpretation of adiabatic (dependent) runs and rewiring based on opinions (Eq.~\eqref{eq:eq2}) can be understood, similar to the opinion models studied by G. I\~niguez et al. ~\cite{inueguez2010} and Friedkin-Johnsen opinion dynamics model and its variants ~\cite{friedkin1990, raineri2025}. Therefore we consider dependent runs where we start with random $\theta$'s at $A=0$ and relay the asymptotic value  of $\theta$'s from this run as the initial condition for the next value of $A$ while slowly varying the limiter $A$ from $0$ to $\pi$ and then backwards from $\pi$ to $0$.  We call these simulations as dependent runs as the initial values of opinions ($\theta$'s) for a particular $A$ are taken from the final values of opinions of the previous run (see Fig.~\ref{fig:fig2}). Again here the latitude of non-commitment is absent ($B-A=0$). As the latitude of acceptance $A$ increases (indicated by a rightward arrow), we see a transition in states from scattered state to $\pi$ state and then an abrupt shift to a consensus state. This forward evolution is referred to as the forward run and is represented by blue circles in Fig. ~\ref{fig:fig5}. The abrupt change in the order parameter and the corresponding macrostate is referred to as a tipping point. Subsequently, we reverse the direction of change in $A$ by gradually decreasing $A$ from $\pi$ to 0 (indicated by a leftward arrow) or the backward run, the order parameter $R$ of the asymptotic state is shown with red triangles in Fig.\ref{fig:fig5}. As can be seen that during the backward run, the latitude of rejection increases while the latitude of acceptance decreases, leading to a different dynamical state for the same value of $A$. In the forward run, the system exhibits a brief scattered state, followed by a broad $\pi$ state, and then transitions into a narrow consensus state. In contrast, during the backward run, the system sustains the consensus state for a larger parameter range, exhibiting multistability between $\pi$ and consensus states over a significant range of $A$. This region is highlighted by a light yellow background, and the system eventually undergoes an abrupt shift from consensus to a fragmented or an incoherent state, marking another tipping point in the system. The blue, pink and yellow background colors in Fig.~\ref{fig:fig5} represent scattered states, consensus and multistable regions respectively. The system's tendency to remain in its current state, despite changes in parameter, is captured by the hysteresis loop formed by the forward and backward variation of the limiter $A$, as can be seen in Fig.~\ref{fig:fig5}. This variation of $R$ with $A$ is demonstrated for different widths, $\gamma$ of the conviction distribution of individuals in the population (see Fig.~\ref{fig:fig5}(a)-(d)). The variation of weighted order parameter, $S$ with $A$ corresponding to Fig.~\ref{fig:fig5} is shown in Fig.~\ref{fig:fig6} where the latency in the system's dynamics and distinction between scattered and $\pi$ states are clearly visible. From Fig.~\ref{fig:fig5} and \ref{fig:fig6}, we observe that for higher $\gamma$, the population fails to reach a strong consensus indicated by decreasing value of $R$ in the consensus region. Also the transition from scattered state to a consensus becomes less abrupt as $\gamma$ increases. Notably the retention of consensus states even for reduced latitude of acceptance may help explain the persistence of assimilation effects under unfavourable conditions~\cite{iatsenko2013}. Furthermore, the area enclosed by the hysteresis loop decreases as the heterogeneity ($\gamma$) of the population increases, indicating diminished system memory and reduced multistability along with reduced order in more diverse populations that have relatively high overall convictions as evident from Fig.~\ref{fig:fig7}. It is analogous to a stubborn or inflexible group with strong conviction tending towards less order and failing to form a consensus~\cite{quante2024}.

\begin{figure}[hptp]
        \centering
        \hspace{-0.2cm}
        \includegraphics[width=0.48\textwidth]{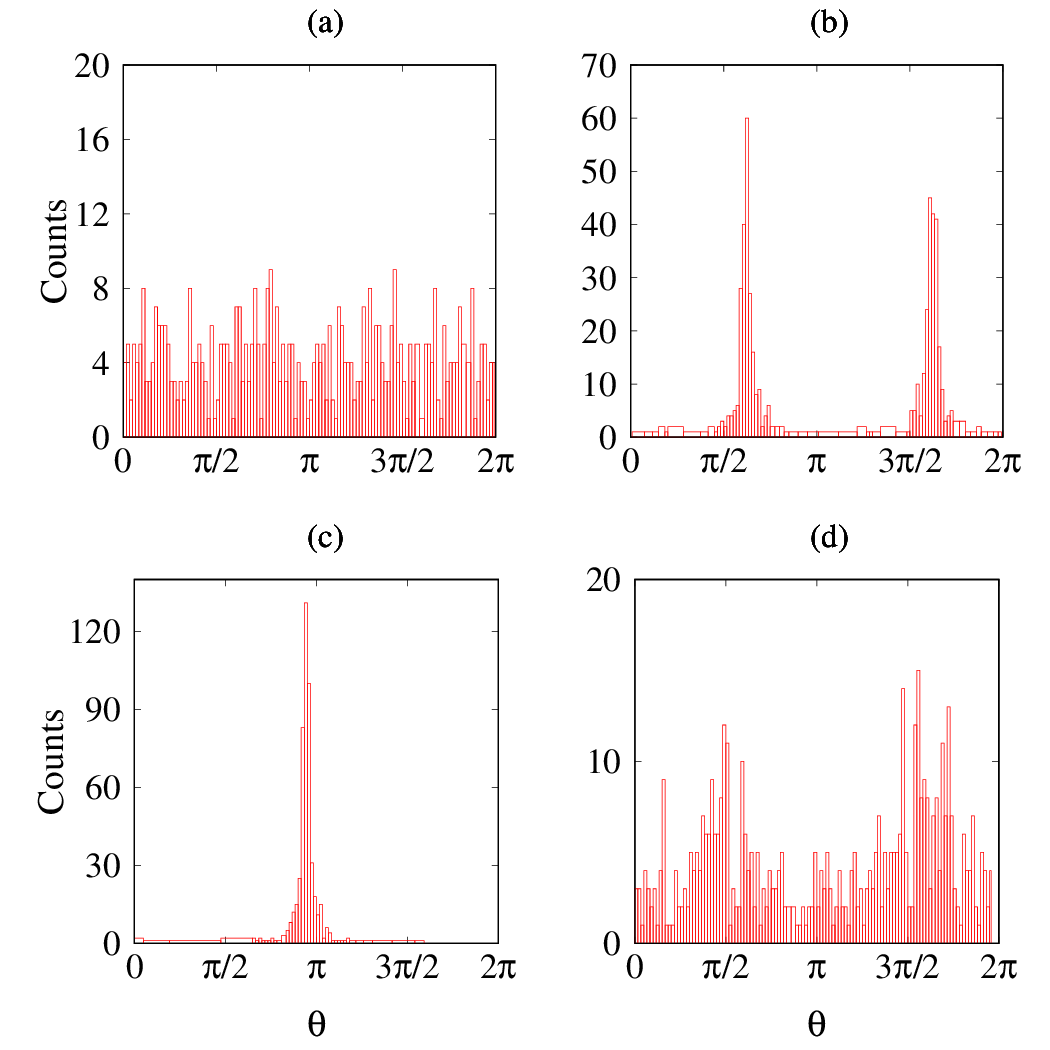} % First image
    
    \caption{Opinion distribution in the (a) scattered, (b) bipolarised, (c) consensus states at $\gamma$=0.05, and (d) bipolarised state at $\gamma$=0.20 from dependent runs shown in Fig.~\ref{fig:fig5}.}
    \label{fig:fig8}
\end{figure}

The distribution of opinions for the observed dynamical states: scattered, $\pi$ (or bipolarised) and consensus is illustrated in Fig.~\ref{fig:fig8}. Fig. ~\ref{fig:fig8}(b) and ~\eqref{fig:fig8}(d) show $\pi$ or bipolarised states for width, $\gamma$ equals $0.05$ and $0.20$ respectively. Comparing these, we observe that the polarised state for $\gamma=0.05$ is more peaked and localized, whereas polarisation for $\gamma=0.20$ is broader and more diffused. Therefore, these plots suggest that when the population is less heterogeneous characterized by low $\gamma$, the population is strongly polarised while for a more heterogeneous population denoted by high $\gamma$ , the opinion polarisation is not strong. For the same reason, greater latitude of acceptance, $A$ will be required to assimilate a highly polarised state into a consensus state as can be seen in increase in light pink background region in Fig.~\ref{fig:fig5}(a)-(d) and Fig.~\ref{fig:fig6}(a)-(d). The $\pi$ and consensus states show multistability for the same values of latitudes of acceptance and rejection i.e $A$, therefore depending upon the initial configuration of the population in the yellow region, the population may get polarised or form a consensus.

\begin{figure}[hptp]

        \centering
        \hspace{-0.2cm}
        \includegraphics[width=0.48\textwidth]{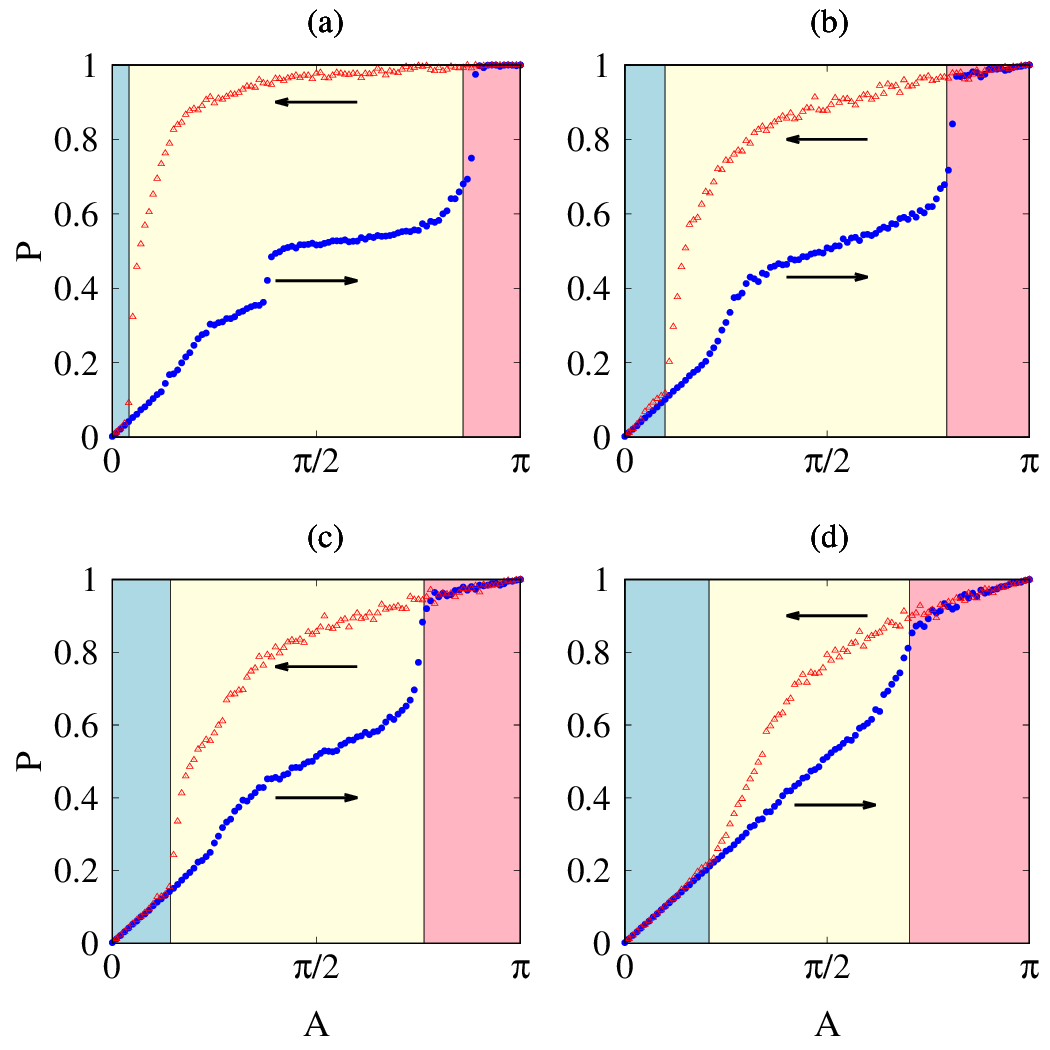} % First image
    
    \caption{Fraction of attractive connections, $P$ with A for forward (blue circles) and backward (red triangles) variation of the limiter for the dependent runs at (a) $\gamma$=0.05, (b) $\gamma$=0.10, (c) $\gamma$=0.15, and (d) $\gamma$=0.20.}
    \label{fig:fig9}
\end{figure}  

\begin{figure}[hptp]

        \centering
        \hspace{-0.2cm}
        \includegraphics[width=0.48\textwidth]{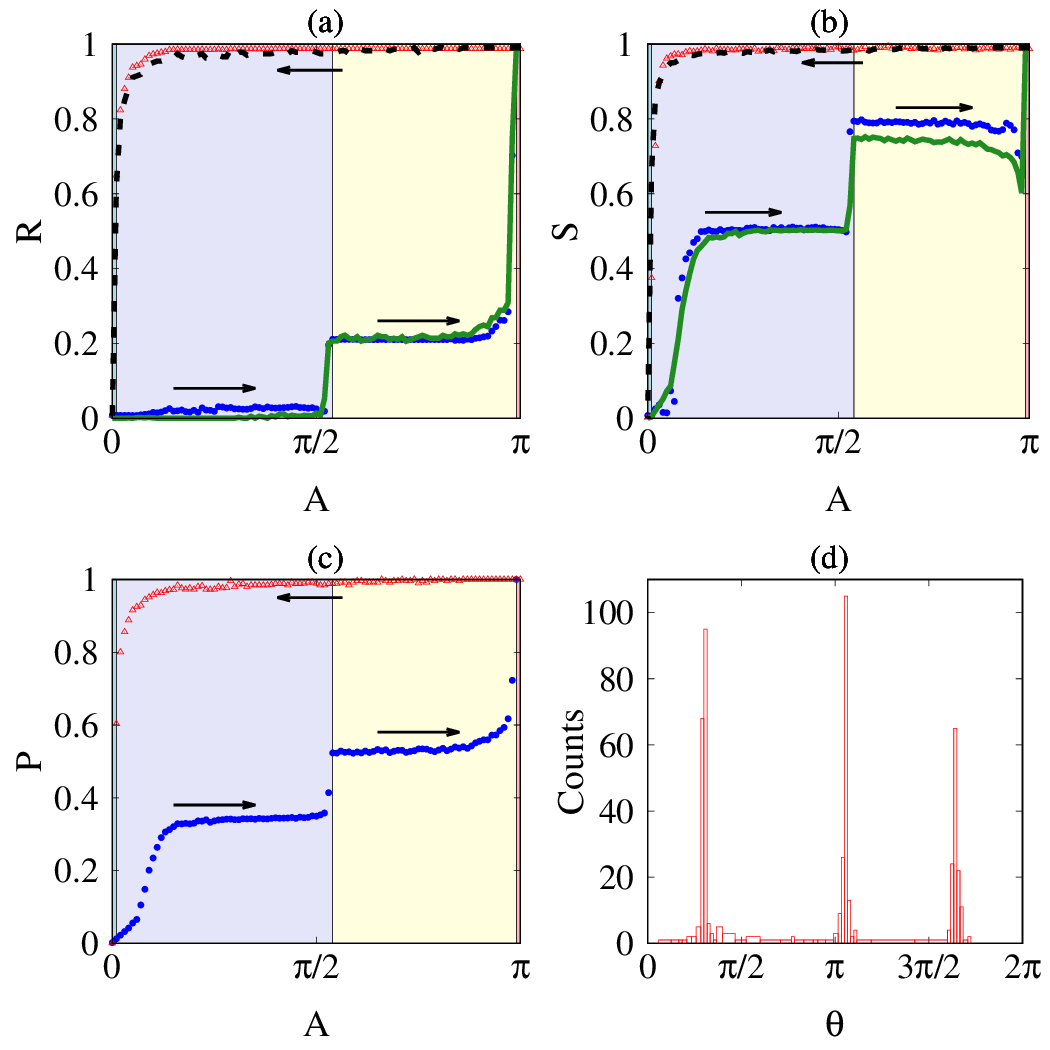} % First image
    
    \caption{(a) Order parameter $R$, (b) average weighted order parameter $S$, (c) fraction of attractive connections $P$ plotted with varying $A$ ($A=B$) for conviction spread $\gamma=0.01$. The numerical results obtained from Eqs.~\eqref{eq:eq1}-~\eqref{eq:eq5} are shown with blue circles in the forward direction and red triangles in the backward direction. The theoretical values (from Eqs.~\eqref{eq:eq19}-~\eqref{eq:eq24} ) are shown with green solid lines (forward run) and black dashed lines (backward run). (d) shows the opinion distribution for the tri-polarised state observed at $A$ = 0.2$\pi$.}
    \label{fig:fig10}
\end{figure}

 In Fig.~\ref{fig:fig9}, the variation in percentage of attractive connections with respect to parameter $A$ is shown, corresponding to the hysteresis behaviour observed in Fig.~\ref{fig:fig5}. A notable slowdown in the rate of change is observed at around 50 $\%$ ($P\approx$ 0.5) mark which corresponds to the $\pi$ state. Although the latitude of acceptance increases linearly with $A$, the percentage of attractive connections does not follow a linear trend in this case. This contrasts with the memory-independent scenario depicted in Fig.~\ref{fig:fig3}, where a more uniform and linear increase in positively coupled oscillators was observed.
  
We observe another interesting feature in this model where further lowering the $\gamma$  (corresponds to populations with weaker conviction or more homogeneity), can induce a two-step transition in the system. The collective state evolves from a scattered configuration to a tri-polarised state and subsequently to a $\pi$ state and a consensus. The position of the emergent clusters depends on the initial conditions, and the clusters may or may not be equally spaced on the phase circle. The evolution of the order parameter $R$, the weighted order parameter $S$, and the fraction of attractive connections $P$ with parameter A are shown in Fig.~\ref{fig:fig10} (a)-(c), while the corresponding phase distribution is presented in Fig.~\ref{fig:fig10} (d). Regions of multistability are indicated by shaded backgrounds, the light purple region denotes coexistence between consensus and tri-polarised or fragmented states, whereas the light yellow region marks coexistence between the $\pi$ states and consensus.
 
 \section{Ott-Antonsen (OA) analysis}

  The system's equation, Eq.~\eqref{eq:eq1} can be rewritten in terms of the weighted order parameter~\cite{ott2008, ott2009} as
  \begin{equation} 
      \frac{d\theta_i}{dt}=v_i=\omega_i + \frac{1}{2i} [W_i e^{-i\theta_i} - \bar{W_i}e^{i\theta_i}],
    \label{eq:eq6}  
  \end{equation}
  where, 
  \begin{equation}
      W_i=\frac{1}{N} \sum_{j=1}^N  K_{ij} e^{i\theta_j}
      \label{eq:eq7}
  \end{equation}
  is the complex weighted order parameter~\cite{hong2012} of the $i^{th}$ agent of the system and $\bar{W_i}$ is its complex conjugate.  
As $N \to \infty$, using ensemble formulation~\cite{barlev2011}, we introduce a probability density function, $f_N(\theta_1, \theta_2,..., \theta_N; \omega_1, \omega_2,..., \omega_N; t)$ for an instant of time $t$. The function follows the oscillator conservation equation,
\begin{equation}
 \frac{\partial f_N}{\partial t} + \sum_{i=1}^N \frac{\partial[f_N \dot{\theta_i}]}{\partial \theta_i} = 0.
 \label{eq:eq8}
\end{equation} 
Therefore in continuum limit, $W_i$ can be written in terms of the distribution function $f_N$ as
\begin{equation}
W_i\equiv \frac{1}{N} \sum_{j=1}^N K_{ij} \int e^{i\theta_j} f_N d^N\omega d^N\theta,
\label{eq:eq9}
\end{equation}
\begin{equation}
W_i= \frac{1}{N} \sum_{j=1}^N K_{ij} \int_{-\infty}^{\infty}d \omega_{j} \int_{0}^{2 \pi} d\theta_{j} e^{i\theta_j} f_j(\theta_j, \omega_j, t)
\label{eq:eq10}
\end{equation}
where $f_j(\theta_j, \omega_j, t)$ is the marginal distribution function~\cite{barlev2011} such that
%\begin{multline}
%f_1(\theta_1, \omega_1, t) = 
%   \int f_N(\theta_2,\theta_3,\ldots,\theta_N;\,
%            \omega_2,\omega_3,\ldots,\omega_N;\,t) \\
%   d\theta_2 \cdots d\theta_N \, d\omega_2 \cdots d\omega_N.
%\label{eq:eq11}
%\end{multline}
\begin{equation}
f_j(\theta_j, \omega_j, t)= \int f_N(\theta_1,\theta_2,\ldots,\theta_N;\,
            \omega_1,\omega_2,\ldots,\omega_N;\,t)\prod\limits_{i \ne j} d\omega_i \, d\theta_i.
\label{eq:eq11}
\end{equation}
Now, multiplying Eq.~\eqref{eq:eq8} by $\prod\limits_{j \ne i} d\omega_j \, d\theta_j$ and integrating, the marginal distribution function satisfies
\begin{equation}
\frac{\partial f_i}{\partial t} + \frac{\partial(f_i \dot{\theta_i})}{\partial \theta}=0.
\label{eq:eq12}
\end{equation}

Expanding $f_i(\theta_i, \omega_i, t)$ as Fourier series gives,
 \begin{equation}
     f_i(\theta_i, \omega_i, t)=\frac{g(\omega_i)}{2\pi}[1+ \sum_{n=1}^\infty(\alpha_i^n(\omega_i, t)e^{in\theta_i} + (\bar{\alpha}_i^n(\omega_i, t)e^{-in\theta_i})]
     \label{eq:eq13}
 \end{equation}
 where $\alpha_i (\omega_i,t )$ denotes the $\theta$ independent coefficients of Fourier series such that $|\alpha_i(\omega_i, t)|<1$ and $\bar{\alpha_i}$ are the complex conjugate of these coefficients.

  Substituting Eqs.~\eqref{eq:eq6} and ~\eqref{eq:eq13} in Eqs. ~\eqref{eq:eq10} and ~\eqref{eq:eq12} we get the evolution equations for the Fourier coefficients as, 
  \begin{equation}
      \dot{\alpha_i}=-i\omega_i\alpha_i + \frac{1}{2}(\bar{W_i}-W_i \alpha_i^2)
      \label{eq:eq14}
  \end{equation}
where, 
\begin{equation}
    W_i=\frac{1}{N}\sum_{j=1}^N K_{ij} \int_{-\infty}^{\infty} \bar{\alpha_j}(\omega_j, t)g(\omega_j)d\omega_j. 
    \label{eq:eq15}  
\end{equation}
After contour integration with respect to $\omega$ and closing the contour in the lower half plane in $\omega$ space,
\begin{equation}
    W_i=\frac{1}{N}\sum_{j=1}^N K_{ij}  \bar{\alpha_j}(-i\gamma, t). 
    \label{eq:eq16}  
\end{equation}
Taking $z_i(t)=\alpha_i(-i\gamma, t)$, Eq.~\eqref{eq:eq14} reduces to

\begin{equation}
    \dot{z_i}(t)=-\gamma z_i + \frac{1}{2}(\bar{W_i}- W_i z_i^2)
    \label{eq:eq17}
\end{equation}
where, 
\begin{equation}
    W_i=\frac{1}{N}\sum_{j=1}^N K_{ij} \bar{z_j}(t).
    \label{eq:eq18}
\end{equation}
Since $K_{ij}$ can attain both positive and negative values, the system can be divided into two sub-populations: one with positive interactions and the other with negative interactions. This is in contrast with the analysis done by Hong and Strogatz ~\cite{hong2011} where the order parameter is expressed in terms of proportions of attractive ($p$) and repulsive ($1-p$) connections since in their model these proportions were same for all oscillators and hence the system can be reduced to a single set of equations (independent of $i$). However, in our case the fractions of attractive ($K_{ij}> 0$) and repulsive ($K_{ij}<0$) connections are different for different agents, therefore we use the ensemble formalism~\cite{barlev2011} that gives us local order parameters ($Z_i$, $W_i$) which in turn can be used to determine the global dynamics of the system ($R$, $S$). Let ${z}_i^p$ and ${z}_i^n$ represent the complex order parameters for the attractively and repulsively connected sub-population respectively. Therefore, Eq.~\eqref{eq:eq17} and Eq.~\eqref{eq:eq18} can be rewritten as:
\begin{equation}
    \dot{z_i}^p(t)=-\gamma z_i^p + \frac{1}{2}(\bar{W_i} - W_i (z_i^p)^2),
    \label{eq:eq19}
\end{equation}

\begin{equation}
    \dot{z_i}^n(t)=-\gamma z_i^n + \frac{1}{2}(\bar{W_i} - W_i (z_i^n)^2),
     \label{eq:eq20}
\end{equation}
and
\begin{equation}
    W_i=\frac{1}{N}\sum_{j=1}^NK_{ij}[\bar{z}_j^p+\bar{z}_j^n]
    \label{eq:eq21}
\end{equation} where,
\begin{equation}
	z_j =
	\begin{cases}
		z_j^p & \text{if, } K_{ij}>0 \\
		z_j^n  & \text{if, } K_{ij}<0 .
	\end{cases}
	\label{eq:22}
  \end{equation}

  The complex order parameter for the $i^{th}$ unit in the system, 
\begin{equation}
     Z_i=\frac{1}{N}\sum_{j=1}^N[\bar{z}_j^p+\bar{z}_j^n].
    \label{eq:eq23}
\end{equation}
Also,
\begin{equation}
R=< |Z_i |>=\frac{1}{N}\sum^N_{i=1} | Z_i |,
\label{eq:eq24}
\end{equation}
and
\begin{equation}
S=<|W_i|> = \frac{1}{N}\sum^N_{i=1}|W_i|.
\label{eq:eq25}
\end{equation}

Note that $R$ and $S$ yield the global order parameter and average weighted order parameter of the system respectively where $<.>$ denotes the average over all the individuals (or oscillators). For theoretical predictions, we fix the limiters ($A$ and $B$) and integrate Eqs.~\eqref{eq:eq19}-\eqref{eq:eq20} where $W_i$ is computed from Eq.~\eqref{eq:eq21} to get steady state values of $z_i^p$ and  $z_i^n$. From these steady state values, the local complex order parameter $Z_i$ is computed using Eq.~\eqref{eq:eq23}. The global order parameters $R$ and $S$ are calculated by averaging the amplitudes of the local complex order parameters as given in Eqs.~\eqref{eq:eq24} and~\eqref{eq:eq25}. For these calculations, the elements of the adjacency matrix $K_{ij}$ are taken from the numerically obtained values generated using algorithm described in Fig.~\ref{fig:fig2} for both dependent and independent runs. From Figs.~(\ref{fig:fig3})-(\ref{fig:fig6}) and Fig.~(\ref{fig:fig10}) we observe that the match between theoretical predictions from OA analysis and the numerical results are good when $\gamma$ is small. As the spread $\gamma$ increases, the numerically computed order parameters deviate from that obtained from OA analysis.

The various emergent dynamical states in the system have been analyzed using Eqs.~\eqref{eq:eq19}-\eqref{eq:eq21}. To probe further into these states we look at the attractive and repulsive parts of the order parameters namely $Z^p_i$ and $Z^n_i$ separately. Therefore when $K_{ij}>0$,
\begin{equation}
    Z^p_i=\frac{1}{N_p}\sum ^{N_p}_{j=1} \bar{z}^p_j 
    \label{eq:eq26}
\end{equation}
and for $K_{ij}<0$,
\begin{equation}
    Z^n_i=\frac{1}{N_n}\sum ^{N_n}_{j=1} \bar{z}^n_j
    \label{eq:eq27}
\end{equation}

where, $N_p$ and $N_n$ denote the number of attractive and repulsive connections for the $i^{th}$ oscillator. Separating the real and imaginary parts of the complex order parameters $Z^p_i$ and $Z^n_i$, we get
\begin{equation}
    R^p_i = |Z^p_i|; \qquad 
    \phi^p_i = \tan^{-1}\!\left( \frac{Im(Z^p_i)}{Re(Z^p_i)} \right)
    \label{eq:eq28}
\end{equation}

\begin{equation}
    R^n_i = |Z^n_i|; \qquad
    \phi^n_i = \tan^{-1}\!\left( \frac{Im(Z^n_i)}{Re(Z^n_i)} \right).
    \label{eq:eq29}
\end{equation}

\begin{figure}[h]

        \centering
       % \hspace{-2.28cm}
        \includegraphics[width=0.50\textwidth]{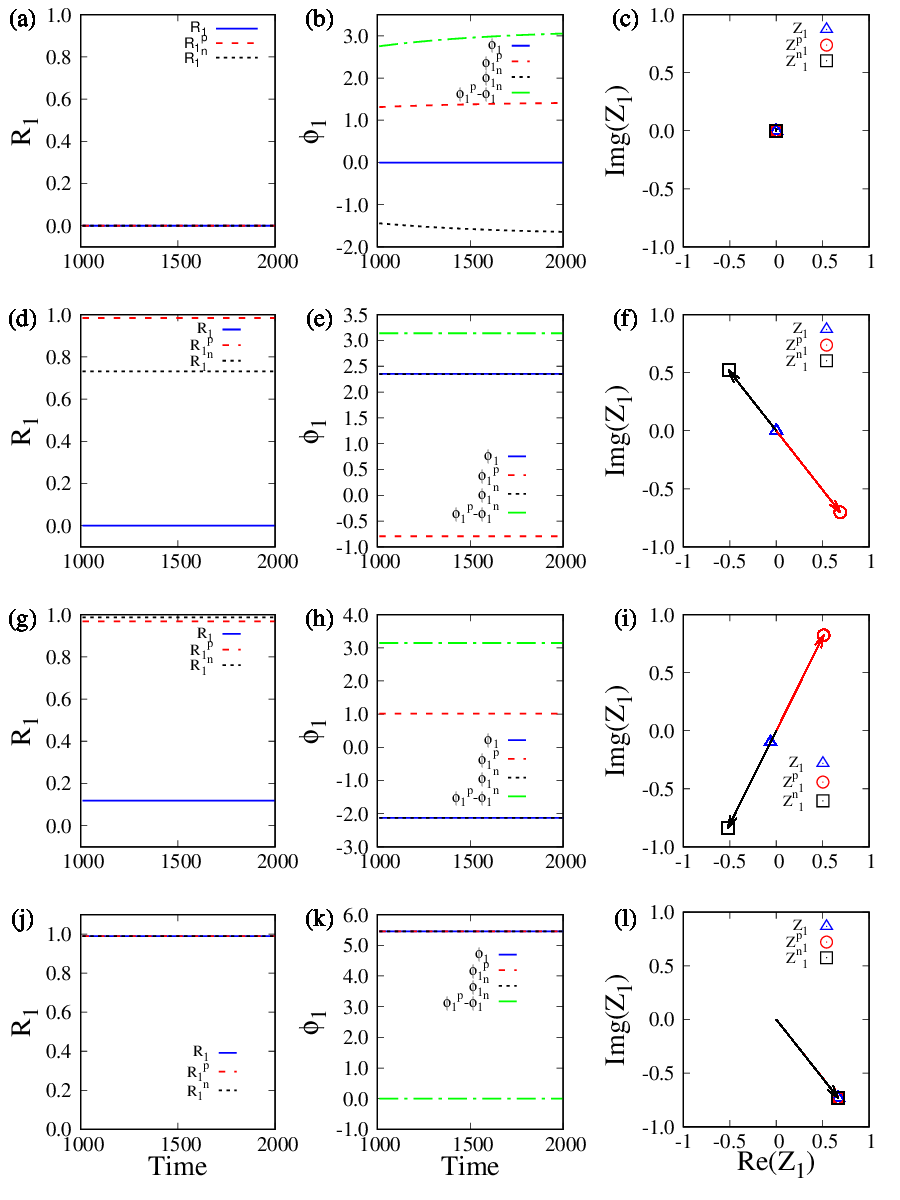} % First image
      \vspace{-1pt}
    \caption{ Components of order parameters computed from Eqs. ~\eqref{eq:eq19}-~\eqref{eq:eq24} and Eqs. ~\eqref{eq:eq26}-~\eqref{eq:eq29} for scattered state (first row (a)-(c)), tri-polarised state(second row (d)-(f)), $\pi$ state (third row (g)-(i)), and consensus state (fourth row (j)-(l)). Time evolution of order parameter amplitudes $R_1$(blue solid line), $R^p_1$(red dashed line), and $R^n_1$(black dashed line) for (a) scattered, (d) tri-polarised, (g) $\pi$, and (j) consensus states respectively are shown. Average phases $\phi_1$(blue solid line), $\phi^p_1$(red dashed line), $\phi^n_1$(black dotted line), and the difference $\phi^p_1 -\phi^n_1$(green dashed-dotted line) are displayed for (b) scattered, (e) tri-polarised, (h)$\pi$, and (k) consensus states. The complex order parameters $Z^p_1$(red $\bigcirc$ symbol), $Z^n_1$(black $\square$ symbol), and $Z_1$(blue $\triangle$ symbol) are plotted for the four states (c) scattered, (f) tri-polarised, (i) $\pi$, and (l) consensus states. The value of $\gamma$ = 0.01 and $A$ for scattered, tri-polarised, $\pi$ and consensus states are taken to be $A=0.01 \pi$, $A=0.2 \pi$, $A=0.7 \pi$, and $A=\pi$ respectively.}
    \label{fig:fig11}
\end{figure}

Eqs.~\eqref{eq:eq28}-~\eqref{eq:eq29} give the order parameters, $R_i^p$ and $R_i^n$ and the corresponding average phases $\phi^p_i$ and $\phi^n_i$ of the positively and the negatively coupled sub-populations with respect to the $i^{th}$ individual. We numerically solve Eqs~\eqref{eq:eq19}-\eqref{eq:eq21} using Runge-Kutta fourth-order method until steady state is reached. Relevant data were extracted from the last 1000 time steps after discarding the transients. For each value of $A$, random initial complex values of $\bar{z}^p_j$ and $\bar{z}^n_j$ are considered and the coupling coefficients $K_{ij}$'s are taken from the numerically observed values of attractive and repulsive connections. Since the individual $R_i^p$ and $R_i^n$ of the oscillators differ depending on the relative attractive and repulsive connections, averaging would obscure important information. Therefore, we analyze the states from the perspective of a single individual and illustrate the results for $i=1$ in Fig.~\ref{fig:fig11}. In the following subsections we describe the various dynamical states using these components of order parameters.

\subsection{Scattered state}
In Fig.~\ref{fig:fig11}(a) we observe that the order parameters of the system with respect to individual-1 $R_1$ (blue solid lines) as well as the order parameters corresponding to the attractively coupled $R^p_1$(red dashed line) and the repulsively coupled $R^n_1$ (black dotted line) sub-populations computed using Eqs.~\eqref{eq:eq26}-~\eqref{eq:eq29} maintain low values close to 0 after the transient period has passed. The average phase $\phi_1$(blue solid line), the average phase of the positively coupled oscillators $\phi^p_1$(red dashed line) and the negatively coupled oscillators $\phi^n_1$(black dotted line) are shown in Fig.~\ref{fig:fig11}(b). $\phi_1$, $\phi^p_1$ and $\phi^n_1$ all reach different values as $\theta$'s are distributed randomly in a scattered state. Hence the difference in phases of attractively and repulsively coupled oscillators $\phi_1^p-\phi_1^n$ is also random as shown by a green dashed-dot line in Fig.~\ref{fig:fig11}(b). The complex order parameters $Z_1$(blue $\triangle$ symbol), $Z_1^p$(red $\bigcirc$ symbol ) and $Z_1^n$(black $\square$ symbol) plotted in the Argand plane are shown in Fig.~\ref{fig:fig11}(c). These complex order parameters lie near the origin indicating lack of coherence in the system. 

\subsection{Tri-polarised state}

As shown in Fig.~\ref{fig:fig10}, the system exhibits multistability between various states, with the emergence of tri-polarised states in the significant range of A. These states are observed in the forward continuation branch, indicated by the right-hand arrow. A closer examination of these states reveals that the sub-population order parameters, $R^p_1$(red dashed line) and $R^n_1$(black dashed line) attain relatively high values (Fig.~\ref{fig:fig11}(d)), though their magnitudes vary across different realisations depending on the specific cluster configuration (equidistant or not). When both sub-populations exhibit strong coherence, the system organises into well-defined equidistant tri-polarised clusters. Furthermore, the overall phase $\phi_1$(blue solid line) tends to align with the dominant sub-population, maintaining a constant phase difference, $\phi^p_1 -\phi^n_1$ (green dashed dotted line) between the attractive and repulsive groups (see Fig.\ref{fig:fig11}(e)). This behavior is also evident in the complex order parameter representation on the Argand plane represented by $Z_1$(blue $\triangle$ symbol), $Z_1^p$ (red $\bigcirc$ symbol ) and $Z_1^n$ (black $\square$ symbol), where the sub-populations form extended arms (red and black) with a relative phase shift of $\pi$, consistent with their high degree of coherence as shown in Fig.~\ref{fig:fig11}(f).

\subsection{$\pi$ state}
 The third row of Fig.~\ref{fig:fig11} shows the complex order parameter with respect to individual-$1$ and its components when the system exhibits bipolarised or $\pi$ states. As shown in Fig.~\ref{fig:fig11}(g), the order parameters of the positively connected, $R_1^p$(red dashed line) and negatively connected $R_1^n$(black dotted line) individuals evolve to high values. Since $R_1$ is a combination of $R_1^p$ and  $R_1^n$, it attains a low value shown by a blue solid line. We observe in Fig.~\ref{fig:fig11}(h) that $\phi_1$(blue solid line), $\phi^p_1$(red dashed line) and $\phi^n_1$(black dotted line) may or may not coincide with each other depending on the clusters, but there is a constant phase difference between $\phi^p_1$ and $\phi^n_1$ of $\pi$ which is shown with a green dashed-dotted line. The post-transient macroscopic phase portrait of the attractively and repulsively coupled sub-populations are diametrically opposite in the Argand plane (see Fig. ~\ref{fig:fig11}(i)). The complex order parameter $Z_1$(blue $\triangle$ symbol) maintains a low value and hence appears at the centre of the Argand plane as can be seen in Fig.~\ref{fig:fig11}(i).

\subsection{Consensus state}
For the consensus state, the amplitude of order parameters $R_1$(blue solid line), $R_1^p$(red dashed line) and $R_1^n$(black dotted line) with respect to unit-$1$ all converge to a value close to unity. Also the average phases $\phi_1$(blue solid line), $\phi^p_1$(red dashed line) and $\phi_1^n$(black dotted line) overlap indicating assimilation of all opinions in the system (see Fig.~\ref{fig:fig11}(k)). Hence, the difference in the average phases $\phi^p_1-\phi_1^n$(green dashed-dotted line) is nearly zero indicating alignment of phases in the consensus state. The complex order parameters $Z_1$(blue $\triangle$ symbol), $Z_1^p$(red $\bigcirc$ symbol ) and $Z_1^n$(black $\square$ symbol) maintain a high value away from zero with a constant average phase as shown in Fig.~\ref{fig:fig11}(l).   

\begin{figure*}[htbt]

        \centering
        \hspace{-0.4cm}
        \includegraphics[width=1.0\textwidth]{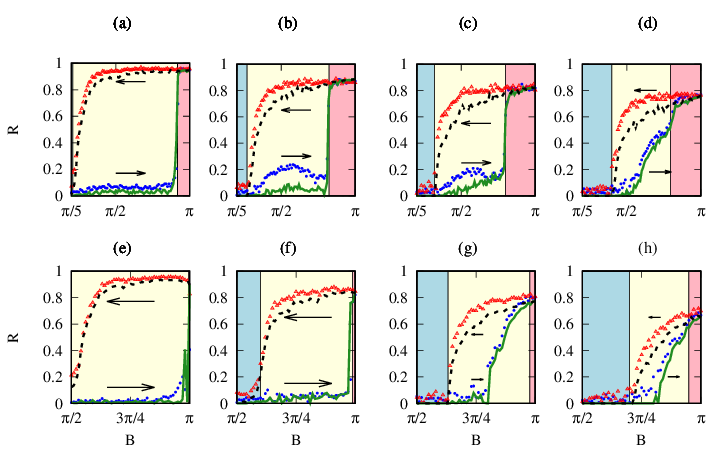} % First image
    
    \caption{Variation of order parameter $R$ with upper limiter $B$, when $B$ increases (forward run) shown with blue circles and when $B$ decreases (backward run) plotted with red triangles in the presence of neutral region. The theoretical values (from Eqs.~\eqref{eq:eq30}-~\eqref{eq:eq36} ) are shown with green solid lines (forward run) and black dashed lines (backward run). The neutral region is characterised by the individuals that do not interact i.e  $K_{ij}=0$.  The range of neutral region $B-A=0.2\pi$ for the sub-figures in the upper panel [(a)-(d)] whereas $B-A=0.5\pi$ for the sub-figures in the lower panel [(e)-(h)]. The width $\gamma$ increases from left to right with $\gamma$=0.05  [(a),(e)] , $\gamma$=0.10  [(b),(f)] , $\gamma$=0.15  [(c),(g)], and $\gamma$=0.20  [(d),(h)].}
     \label{fig:fig12}
\end{figure*}

\begin{figure*}[htbt]

        \centering
        \hspace{-0.4cm}
        \includegraphics[width=1.0\textwidth]{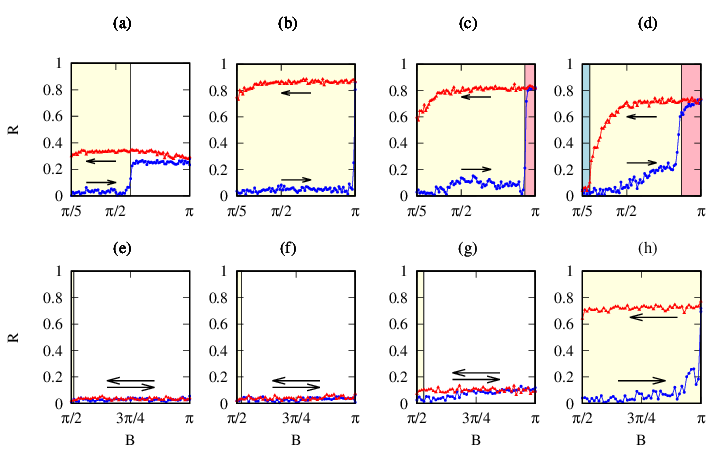} % First image
    
    \caption{Order parameter $R$ for forward (blue circles) and backward (red triangles) variation of the upper limiter $B$ taking the condition $K_{ij}(B \pm \delta B)=K_{ij} (B)$ for the opinion pairs falling in the neutral region with different values of distribution width $\gamma$=0.05  [(a), (e)], $\gamma$=0.10 [(b), (f)], $\gamma$=0.15  [(c), (g)], and $\gamma$=0.20 [(d), (h)]. The sub-figures in the upper panel [(a)-(d)] are for neural range $B-A=0.2\pi$ and the sub-figures in the lower panel [(e)-(h)] are for $B-A=0.5\pi$.}
     \label{fig:fig13}
\end{figure*}
  
 \section{Effect of neutral region} 

\subsection{Non-interacting neutral attitudes (if $A<|\theta_j-\theta_i|<B$, then $K_{ij}=0$)}
In a population the individuals can attain attractive attitude ($K_1$) if their opinions lie in the latitude of acceptance (when $|\theta_j -\theta_i|<A$) or repulsive attitude ($K_2$) if the opinions belong to the latitude of rejection (when $|\theta_j -\theta_i|>B$), in addition there could also be interactions that are neither positive nor negative in attitude. To incorporate such interactions we introduce latitude of non-commitment in our model such that the differences in opinions lying in the range $A< |\theta_j -\theta_i|<B$ where $B-A \neq 0$ ($A<B$ ) fall in the latitude of non-commitment shown with yellow region in Fig.~\ref{fig:fig1}(b). In this case, pair of individuals for which phase differences lie in this neutral range does not interact i.e $K_{ij}=0$ (Eq.~\eqref{eq:eq2}). Since $K_{ij}$ can now attain positive ($K_1$), negative ($K_2$) and zero values, Eqs.~\eqref{eq:eq17} and~\eqref{eq:eq18} are divided into three equations corresponding to the three sub-populations wherein the individuals have attractive, repulsive and neutral interactions. Let $z_i^p$, $z_i^n$ and $z_i^{neu}$ represent the complex variables for attractively, repulsively and neutrally coupled sub-population respectively. Therefore Eqs.~\eqref{eq:eq17} and \eqref{eq:eq18} can be rewritten as:
\begin{equation}
    \dot{z_i}^p(t)=-\gamma z_i^p + \frac{1}{2}(\bar{W_i} - W_i (z_i^p)^2),
    \label{eq:eq30}
\end{equation}

\begin{equation}
    \dot{z_i}^n(t)=-\gamma z_i^n + \frac{1}{2}(\bar{W_i} - W_i (z_i^n)^2),
     \label{eq:eq31}
\end{equation}

\begin{equation}
    \dot{z_i}^{neu}(t)=-\gamma z_i^{neu} + \frac{1}{2}(\bar{W_i} - W_i (z_i^{neu})^2),
     \label{eq:eq32}
\end{equation}
and
\begin{equation}
    W_i=\frac{1}{N}\sum_{j=1}^NK_{ij}[\bar{z}_j^p+\bar{z}_j^n+\bar{z}_j^{neu}]
    \label{eq:eq33}
\end{equation} where,
\begin{equation}
	z_j =
	\begin{cases}
		z_j^p & \text{if, } K_{ij}>0 \\
		z_j^n  & \text{if, } K_{ij}<0 \\
		z_j^{neu} & \text{if, } K_{ij}=0.
	\end{cases}
	\label{eq:34}
  \end{equation}

  The complex order parameter for the $i^{th}$ unit in the system, 
\begin{equation}
     Z_i=\frac{1}{N}\sum_{j=1}^N[\bar{z}_j^p+\bar{z}_j^n+\bar{z}_j^{neu}].
    \label{eq:eq35}
\end{equation}
Also,
\begin{equation}
R=<|Z_i|> = \frac{1}{N}\sum^N_{i=1}|Z_i|,
\label{eq:eq36}
\end{equation}

Eq.~\eqref{eq:eq36} gives the order parameter of the system for a particular value of $A$ and $B$, $K_{ij}$ is taken from the simulation of Eqs.~\eqref{eq:eq1}-\eqref{eq:eq2}. We fix the region of non-commitment and increase the limiters $B$ (or $A$ ) to their full extent in the forward direction and then backwards. We consider the dependent runs in order to capture the potential tipping points and hysteresis loops. The upper panel in Fig.~\ref{fig:fig12} presents the results for $B-A=0.2\pi$ meaning out of the total range of phase differences, 20 $\%$ is occupied by the neutral region. We start with the parameters $A=0$ and $B=0.2\pi$ and increment $A$ and $B$ maintaining the neutral region at $ B-A=0.2\pi$ in the forward direction (indicated by blue circles and arrow pointing towards right) and then in the backward direction by decreasing $B$ (indicated by red triangles and arrow towards left) and plot the order parameter $R$ (see Fig.~\ref{fig:fig12}). In Fig.~\ref{fig:fig12}, the theoretical $R$ values obtained from Eqs.~\eqref{eq:eq30}-~\eqref{eq:eq36} are shown with green solid lines for forward runs and black dashed lines for backward runs. From Fig.~\ref{fig:fig12}, we observe that in the forward run (blue circles), the states change from scattered to $\pi$ to a consensus with an explosive transition from $\pi$ state to a consensus state. The point of transition from $\pi$ state to consensus shifts to a lower value of $B$ as the width $\gamma$ increases implying that a population with broader conviction distribution is more susceptible to change in opinions and forming consensus with respect to increase in the latitude of rejection (or acceptance), compared to a narrowly peaked distribution where a less diverse population tend to stick with their opinions and fail to arrive a common decision even if there are significant changes in the environment~\cite{schawe2021}. In the backward run, the state maintains a consensus until another tipping point is reached, beyond which the state changes from a consensus to a fragmented state. The presence of two tipping points, one in the forward run and the other in the backward run along with overlap of initial and final macroscopic measure indicates that the process is reversible~\citep{macy2021}. If the neutral region is increased to $B-A=0.5\pi$, again two tipping points are observed indicating reversibility of state as shown in the lower panel of Fig.~\ref{fig:fig12}(f)-(h). The degree of coherence in the consensus region seems independent of the range of the neutral region meaning for a particular $\gamma$, $R$ values in the consensus region are nearly equal in the upper and lower panels of Fig.~\ref{fig:fig12}. Also, for both 0.2 $\pi$ and 0.5 $\pi$ neutral ranges (upper and lower panels), the area under the hysteresis loop (yellow region) decreases with increase in $\gamma$. For the same $\gamma$, the system exhibits scattered states for a larger range of $B$ when the neutral region is large while the range for consensus diminishes. This implies that the undecided individuals (denoted by the neutral region $K_{ij}=0$) in a population favours the scattered opinions and are comparatively less favourable to consensus formation~\citep{balenzuela2015}.

\subsection{Stubborn neutral attitudes (if $A<|\theta_j-\theta_i|<B$, then $K_{ij} (B+\delta B) = K_{ij} (B)$)}

From the above results it is evident that the attitude of the individuals whose opinions lie in the neutral region can drastically affect the overall state formation~\citep{balenzuela2015,wu2023} in the system. Therefore, we ask that instead of having non-interacting individuals ($K_{ij} =0$) in the neutral region what will happen if the neutral region is characterised by the presence of stubborn individuals that do not change their opinions even when there are changes in the social environment \ie $K_{ij} (B) = K_{ij} (B+\delta B)$. Therefore, the individual's attitude in neutral region retains the previously made connections and do not change with change in either $A$ or $B$. The dynamical states for this case are captured in Fig.~\ref{fig:fig13}. When the neutral region $B-A=0.2\pi$ (upper panel), for low $\gamma$ the state changes from a scattered state to a tri-polarised state to a $\pi$ state in forward run (see Fig.~\ref{fig:fig13}(a)). When $B$ decreases (backward run), the system remains in the $\pi$ state throughout. Hence in this case there is a multistability between tri-polarised and bipolarised states and also between bipolarised ($\pi$ state) and scattered states. In addition, the process has now become irreversible as the system stays in $\pi$ state even on fully reverting the parameters. For higher $\gamma$, we observe a change in the nature of multistability wherein the consensus and $\pi$ state or consensus and scattered states coexist with irreversibility as shown in Figs. ~\ref{fig:fig13}(b)-(c). When $\gamma$ is increased further, the transitions become reversible with the presence of a hysteresis region where consensus states coexist with the scattered states and $\pi$ states as can be seen in Fig. ~\ref{fig:fig13}(d). If we consider a larger neutral region, say $B-A=0.5\pi$, we observe scattered and $\pi$ states for smaller values of $\gamma$ as shown in the lower panel of Fig.~\ref{fig:fig13}. During the forward run, the system goes from a scattered state to a $\pi$ state and then remains there even on increasing the latitude of acceptance to its maximum limit ($B$=$\pi$). The backward run is also dominated by the presence of $\pi$ states and there is a very small region of multistability where scattered states and $\pi$ states coexist (see Figs.~\ref{fig:fig13}(e)-(g)). These results also confirm that the population with a significant number of undecided or stubborn individuals can lead to a situation of not being able to unite, even for a common cause~\cite{macy2021}. Studies on Iranian revolution suggest that even when there was full communication ($K_{ij} \neq 0$), the individuals failed to form a consensus of opinions because of their social and cognitive limitations that prevented them to change their attitudes and hence retaining their previous attitudes ($K_{ij} (B+\delta B) = K_{ij} (B)$) towards a topic~\cite{kuran1991} which can also be seen in Fig.~\ref{fig:fig13}(a), (e)-(g). The effect of diversity can be observed in Fig.~\ref{fig:fig13}(c) and (h) where consensus of opinions does occur abruptly which is analogous to an irreversible revolution~\cite{bordonaba-bosque2025, thomassen2012}. More diversity results in the abrupt change being reversible as can be seen in Fig.~\ref{fig:fig13}(d) which is similar to restorations after the revolution has occurred~\cite{norkus2022}. There are instances where changing the attitude from ignorant to stubborn can be advantageous, for instance in western ghats of India certain groves were considered to be sacred, the ignorance about this religious association leads to deforestation of the groves whereas changing the individual attitudes from ignorant ($K_{ij}=0$) (Fig.~\ref{fig:fig12}) to conservative ($K_{ij} (B+\delta B) = K_{ij} (B)$) (Fig.~\ref{fig:fig13}) prevented the consensus to cut these groves, thereby saving them ~\cite{gadgil1976}. When the neutral region is large then even a diverse population with high $\gamma$ shows irreversible multistability between states as can be seen in Fig.~\ref{fig:fig13}(h) where there is a non-reversible multistability between the consensus and $\pi$ states and also consensus and scattered states. We note that the theoretical predictions using OA analysis (Eqs.~\eqref{eq:eq30}-~\eqref{eq:eq36}) for stubborn neutral attitude fail to predict the exact points of transitions and order parameter values when the neutral range is wider and transitions are irreversible. Therefore, for this case we show only the numerically obtained results in Fig.~\ref{fig:fig13}. However qualitatively the OA analysis predicts similar states and state transitions as observed in numerical simulations.

\begin{figure}[htbt]
        \centering
        \hspace{-0.5cm}
        \includegraphics[width=0.48\textwidth]{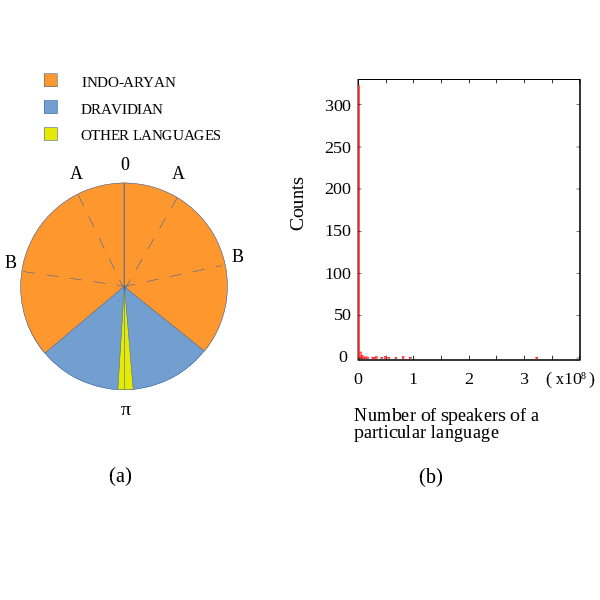} % First image
    
    \caption{(a) Pictorial representation of Indian languages on a phase circle based on their origin and similarity with each other with lower and upper thresholds $A$ and $B$ respectively. (b) Distribution of number of speakers of a particular mother tongue language.}
    \label{fig:fig14}
\end{figure}

\begin{figure}[htbt]
        \centering
        \hspace{-0.2cm}
        \includegraphics[width=0.48\textwidth]{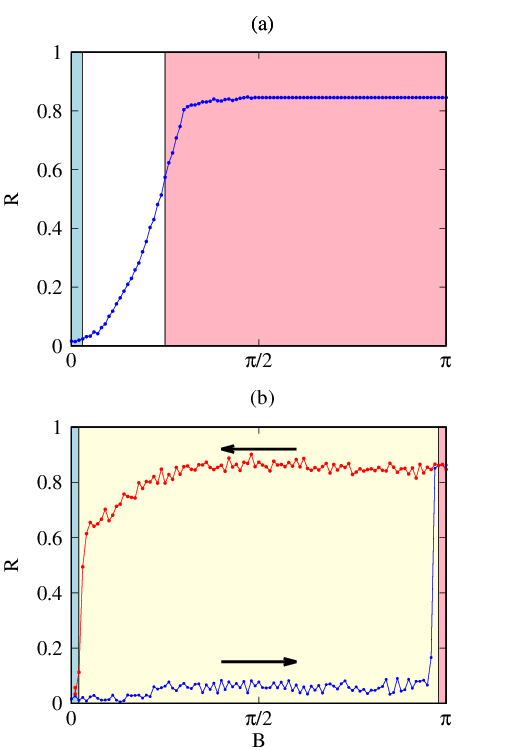} % First image
    
    \caption{Order parameter $R$ is plotted with the change in limiter value, $B$ ($A=B$) for (a) independent runs and (b) dependent runs implementing the model on the language data. For the simulations, we consider $N$=354, Q=1.0, $K_1$=0.01, and $K_2$=-0.01.}
    \label{fig:fig15}
    %\vspace{-40pt}
\end{figure}

\begin{figure*}[htbt]

        \centering
        \hspace{-0.4cm}
        \includegraphics[width=1.0\textwidth]{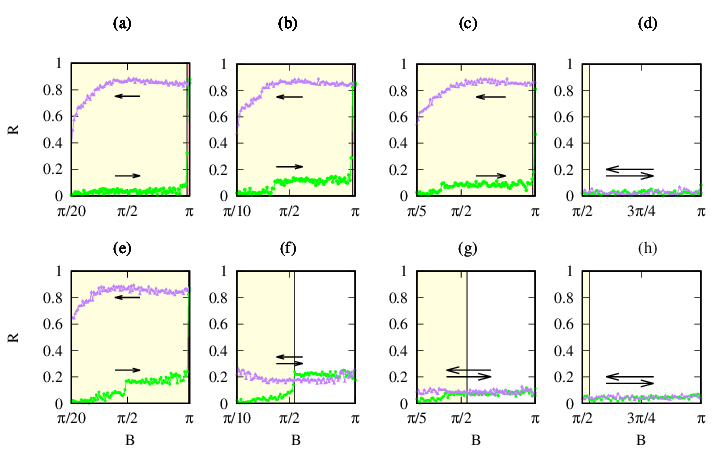} % First image
    
    \caption{Order parameter $R$ with upper limiter, B for forward(green) and backward(purple) variation of limiters. The plots in the top panel [(a),(b), (c), (d)]  and the sub-figures in the lower panel [(e), (f), (g), (h)] are for neutral region $B-A=0.05\pi$, $B-A=0.1\pi$, $B-A=0.2\pi$, $B-A=0.5\pi$ respectively.  The conditions for the neutral region for [(a)-(d)] are $K_{ij}=0$ and for [(e)-(h)] are $K_{ij}(B \pm \delta B) =  K_{ij}(B)$. Other parameter values are $N$=354, Q=1.0, $K_1$=0.01, and $K_2$=-0.01.}
    \label{fig:fig16}
\end{figure*}

\section{Indian languages}

To test the scope of the results obtained from the proposed model, we apply our model to real world data and for this we take the data representing the diversity of languages in India. There is a large pool of languages in India that have competed with each other leading to endangerment or extinction of some languages. There have been observations of language death~\cite{abrams2003, crystal2002} and assimilation of languages~\cite{tran2010} towards a common tongue due to various factors~\cite{crystal2002}. Based on their origin and their proximity to other languages,  these languages are placed on the cyclic scale as shown in Fig.~\ref{fig:fig14}(a). The major portion of languages in India are Indo-Aryan and its variants (orange region), other major languages comprise of the Dravidian language (blue region) sharing a common region since centuries though having a different origin. Other languages (yellow region) consist of Sino-Tibetian, Austro-Asiatic and Tai-Kadai languages comprising only 5 percent of the total speakers in India and having a distant origin as compared to Indo-Aryan languages and hence placed farthest in Fig.~\ref{fig:fig14}(a) with respect to Indo-Aryan languages. We have used number of mother tongue speakers data for 354 Indian languages from census of India (2011)\cite{census2011}, and the frequencies $\omega's$ are computed by normalising them with the maximum value in the dataset, the distribution of which is shown in the histogram plotted in Fig. ~\ref{fig:fig14}(b). The number of speakers of a specific mother tongue language is comparable to the natural frequency, $\omega$ of an oscillator or language in this context, which determines the extent to which it gets affected by other languages. We assign initial $\theta$ values to 354 languages according to Fig.~\ref{fig:fig14}(a) and let the system interact considering attractive coupling as the tendency of a language to assimilate towards a common language. The limiter range $B$ covers languages with common origin initially as can be seen in Fig.~\ref{fig:fig14}(a). Same initial distribution of $\theta$'s has been used for different limiter values to observe a change in state from scattered to $\pi$ to a consensus state for the independent runs shown in Fig.~\ref{fig:fig15}(a). On increasing the range of attractive interactions, $B$ and retaining the phases from the previous run (dependent runs), we observe multistability highlighted by light yellow background in Fig.~\ref{fig:fig15}(b) and the transitions are reversible.

We also introduce the neutral region with attitude $K_{ij}=0$ in the analysis of language data and observe the changes in the states with variation of limiters $A$ and $B$ keeping the neutral region fixed. Fig.~\ref{fig:fig16} shows results for 5, 10, 20, 50 percent(Fig.~\ref{fig:fig16}(a)-(d)) of total range falling in the latitude of non-commitment with ignorant attitude $K_{ij}=0$ and Fig.~\ref{fig:fig16}(e)-(h) with neutral attitudes as $K_{ij}(B\pm \delta B)=K_{ij}(B)$.

Fig.~\ref{fig:fig16}(a), (b), (e), (f) also highlight how the choice of attitudes of the speakers of a particular language within the neutral region can significantly affect the language dynamics as a whole. When the choice of neutral attitude is $K_{ij}=0$, the system attains all the previously attained states with multistability and tipping points between $\pi$ states and consensus states as depicted in Fig.~\ref{fig:fig16}(a)-(b) indicating assimilation towards a common language. Also the states are irreversible once it reaches a consensus evident by the absence of a tipping point in the backward run. When the attitude of interactions in the neutral region is ignorant ($K_{ij}=0$), irreversible consensus state is reached meaning speakers converge towards a common language leading to extinction of other languages~\citep{crystal2002}. On the contrary, if the neutral individuals maintain their previously attained attitudes ($K_{ij}(B \pm \delta B)=K_{ij}(B)$) and for moderate neutral range, they do not reach a consensus meaning never assimilating~\cite{abrams2003} even for large changes in the environment (represented by $B$) as observed in Fig. ~\ref{fig:fig16}(f). Such conservation through change in attitude can be used to revitalise an endangered language through various means\citep{fishman}.

Comparing Fig.~\ref{fig:fig16}(c) and  Fig.~\ref{fig:fig16}(g) we observe that consensus can be achieved or avoided depending on the chosen attitude within the neutral region. However, when the neutral region spans half of the total range, the choice of attitude does not help reaching a consensus as can be seen in Figs.~\ref{fig:fig16}(d) and (h). These findings have strong sociolinguistic implications. While the assimilation of multiple languages into a single dominant language may promote uniformity, it also risks the erosion of linguistic diversity and cultural heritage ~\cite{abrams2003}~\cite{crystal2002}. Our results show that the choice of attitudes of the neutrally coupled native speakers can both hinder or facilitate the assimilation of languages. The backward run(purple line) in Figs. ~\ref{fig:fig16}(a)-(c),(e) is analogous to fragmentation of a single language into many languages due to increasing range of rejection or decreasing $B$ producing variants with a common source, Indo -Aaryan in this case. Thus, this model exhibiting a variety of dynamical states that emerge from different combinations of attractive, neutral, and repulsive interactions, along with the extent of agent's attitudes, offers a compelling framework for understanding social dynamics and opinion formation in different social contexts.

\section{POPULATION TENDENCIES AND STATES}
To understand the scattered, bipolarised ($\pi$ state), and consensus states as behavioural tendencies of a group of individuals,
simulation based study of the model described by Eq.~\eqref{eq:eq1}-~\eqref{eq:eq3} has been done and tendencies of a crowd based on the extent of attitudes has been studied. Since the change in opinions depends on the type of attitude one has towards others, we have considered different types of populations by fixing $A$ and $B$ values and looked at the emergent states as the attractive and repulsive coupling strengths $K_1$ and $K_2$ are varied. 

\begin{figure}[h]

        \centering
        \hspace{-0.7cm}
        \includegraphics[width=0.48\textwidth]{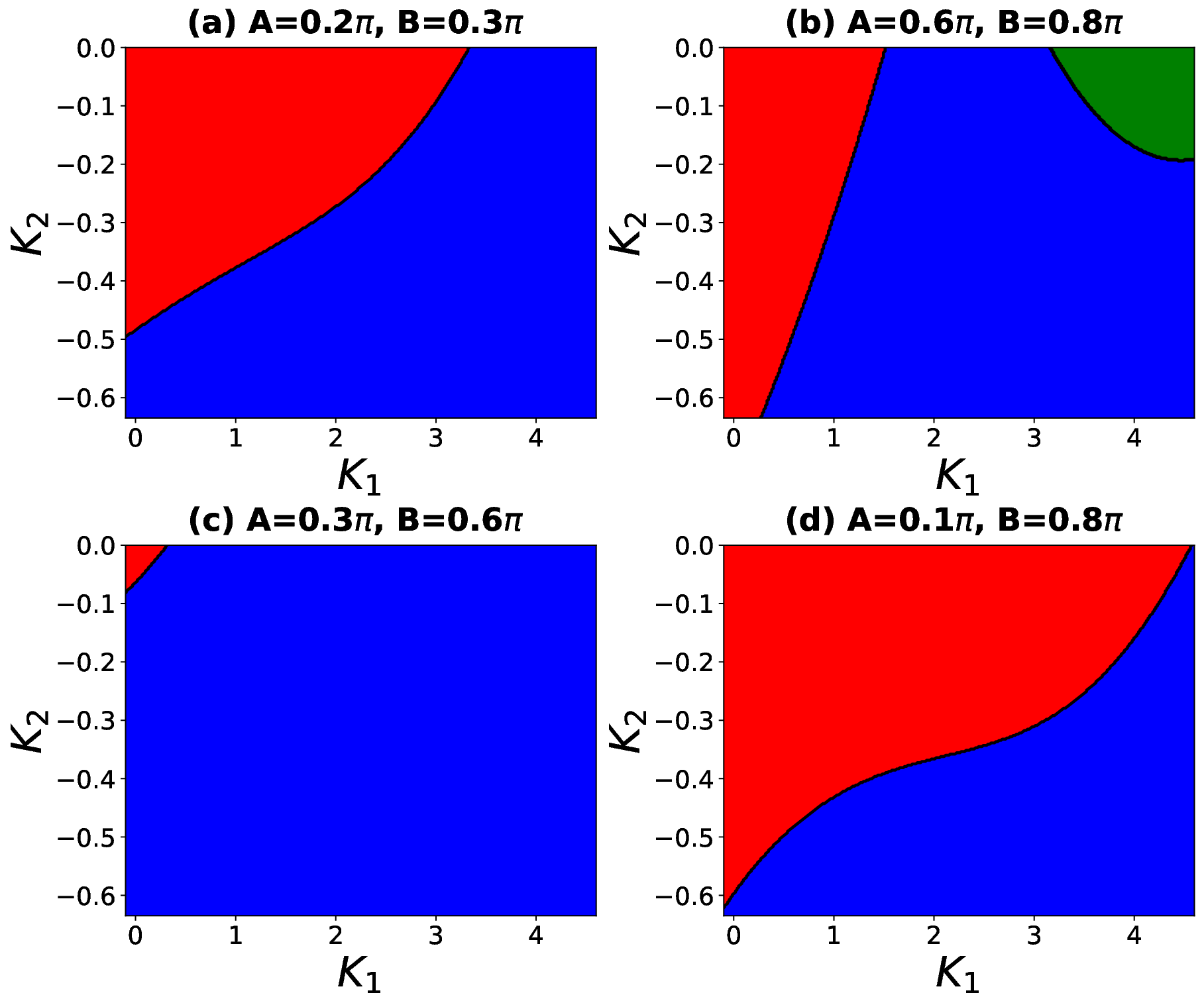} % First image
    
    \caption{Figure showing the existence of scattered(red), $\pi$(blue), and consensus(green) states in the parameter space  $K_1$--$K_2$. The states are obtained for independent run for N=500, $\gamma$=0.05, and taking $K_{ij}=0$ for neutral region.}
     \label{fig:fig17}
\end{figure} 

We consider a case where there is a low latitude of acceptance and non-commitment, high latitude of rejection and high ego evolvement~\cite{jager2005, fisher1999}. To incorporate the above features, we set $A=0.2\pi$ and $B=0.3\pi$ and differences between each pair of oscillators are calculated. Therefore, 20 $\%$ of the total range($\pi$) comprises the latitude of acceptance, 10 $\%$ ($B-A=0.1\pi$) of the total range represents the latitude of non-commitment and 70 $\%$ remaining pair of differences fall in the latitude of rejection amplifying contrast effects. Any interaction in such a population will have a high probability of falling in the latitude of rejection resulting in more repulsive attitudes in the group and one would expect scattered and $\pi$ states  emerging in such populations. The predictions of our model are in accordance with these observations as can be seen in Fig.~\ref{fig:fig17}(a). Random initial phases have been taken for each $K_1-K_2$ value and the asymptotic state is detected by analysing the phase and frequency distribution of the steady state. Fig.~\ref{eq:eq17} shows different states formation for different extents of attitudes in terms of attractive(positive) and repulsive(negative) couplings. The attitudes of the neutrally coupled individuals are considered to be non-interacting ($K_{ij}=0$). The neutral attitude $K_{ij} = K_{ij}$ is redundant in this case since we are looking at the states at fixed thresholds $A$ and $B$.

For the case where $A=0.6\pi$ and $B=0.8\pi$, such that the group has high latitudes of acceptance and lower latitude of rejection and non-commitment, we observe scattered(red), $\pi$(blue) and consensus(green) states in different regions of parameter $K_1$-$K_2$ (see  Fig.~\ref{fig:fig17}(b)). Lower values of attractive coupling $K_1$ result in scattered states, intermediate values of $K_1$ lead to $\pi$ states whereas higher $K_1$ and low $K_2$ values give rise to a consensus.  When there is a small latitude of acceptance ($A=0.3\pi$) and non-commitment and greater latitude of rejection, $\pi$ state is sustained for most coupling regions (Fig. ~\ref{fig:fig17}(c)). For the case shown in ~\eqref{fig:fig17}(d), we consider a population with very small latitude of acceptance($A=0.1\pi$) and high latitude of non commitment ($B=0.8\pi$) which on varying attitudes ($K_1$ and $K_2$), lead to scattered and $\pi$ states only. Thus, we observe various behavioural tendencies in the opinion dynamics for different population types that are characterised by different degrees of attitudes ($A$ and $B$). From these observations we can conclude that the latitude of acceptance favours consensus formation while the latitude of rejection is favourable to polarisation or scattered states. The presence of neutral connections can affect the collective dynamics in such a way that the non-interacting undecided individuals ($K_{ij}$ = 0), can make a polarized population less polarized but with scattered opinions.

\section{Conclusion}

We investigate the emergent dynamics of opinions in a population of interacting individuals using a variant of KM where the interactions can be attractive, repulsive or neutral based on the differences in opinions of the interacting agents. As the range of attractive and repulsive coupling is varied, the system transitions from scattered to clustered and then to consensus states. The clustered states can be bipolarised ($\pi$ state) or tri-polarised. The transitions from tri-polarised to bi-polarised and from bipolarised to consensus states are explosive in nature accompanied by hysteresis. The existence of bipolarised and tri-polarised states with symmetric pairwise interaction is interesting and is in contrast with the dynamical states observed in the case of asymmetric pairwise coupling~\citep{hong2012}. We demonstrate tipping points, multistability and hysteresis loop with respect to the range of coupling attitudes. The multistability also explains the existence of consensus of opinions even for a small reach of attractive coupling depending on the initial configuration of the system. Our results indicate that the attitude (nature of interaction) of undecided or stubborn individuals can be leveraged to understand and manipulate polarisation. Increasing heterogeneity among the individuals tends to reduce the region of multistability and the population struggles to reach a strong consensus. These results can provide useful insights for devising strategies to mitigate polarisation or radicalisation (strong consensus) in a group by tuning system's parameters such as attitudes and interaction ranges. Also, prior information of tipping points of abrupt transitions can facilitate in avoiding undesirable states in different social contexts. 
The model is relevant for a variety of sociophysical settings of opinion formation because of its tunability. To explore real world relevance, we incorporate empirical data on the number of mother tongue speakers, treating it as a proxy for the natural frequencies (conviction) in the population. This allows us to model the dynamics of assimilation of multiple languages into a dominant language. The preservation, assimilation, reversibility of language shift can significantly depend on the attitude in the neutral region as well as on the heterogeneity or width of the distribution of mother tongue speakers. Also the model has been implemented on populations of different types and tendencies showing how change in attitude of neutrally behaving individuals can bring about drastic changes in the cumulative opinions, tipping points, multistability, hysteresis loops and reversibility of states. The understanding gained from this model study can be applied to harness and control the polarisation of opinions during critical periods, as there are empirical evidences from experiments~\citep{voelkel2024} and simulations~\citep{mitigating2025} demonstrating that opinions can indeed shift over time under proper control. We have not considered the presence of bilingual or multilingual individuals in the system, which can be one of the future directions to explore. Furthermore, the effect of higher-order interactions may be incorporated to capture the influence of multiple individuals beyond pairwise interactions~\citep{skardal2020}. Appropriate scaling of the model could also be explored to enable its qualitative application to real-world scenarios.

\begin{acknowledgments}
 SP gratefully acknowledges the UGC fellowship for financial support. SRU is thankful to BHU for providing financial assistance through seed grant under the IoE scheme. The authors acknowledge the National Supercomputing Mission for the computational facility PARAM Shivay at IIT BHU.

\end{acknowledgments}

\section*{Data Availability}
 The dataset for the number of mother tongue speakers analysed during the study is available online, the link for which is provided in the references.

\section*{Conflict of interest}
 The authors have no conflicts of interest to declare.

%\bibliography{aipsamp}

\end{document}